\documentclass[traditabstract]{aa} 

\usepackage[]{url}
\PassOptionsToPackage{hyphens}{url}

\usepackage[varg]{txfonts}
\usepackage{graphicx, times, float, rotating, color, lscape, bm, mathtools}
\usepackage{amssymb, amsmath}
\usepackage{caption, subcaption, mwe, gensymb, booktabs, dirtytalk, lmodern, fancyhdr, tabularx}
\usepackage{wrapfig, floatflt, natbib}
\usepackage{placeins}
\usepackage[version=4]{mhchem}
\usepackage{multicol}

\newcommand{\msol}{\mbox{M$_{\odot}$}} 
 
\newcommand{\msolyr}{{M$_{\odot}$}\,yr$^{-1}$} 
\newcommand{\mdot}{$\dot{M}$}
\newcommand{\lsol}{\mbox{L$_{\odot}$}}

\newcommand{\jks}{Jy km s$^{-1}$} 
\newcommand{\kms}{km s$^{-1}$} 
\newcommand{\ks}{km s$^{-1}$}

\newcommand{\fco}{$f_{\rm CO/H_2}$}

\AtBeginDocument{%
  \setlength{\abovedisplayskip}{6pt}%
  \setlength{\belowdisplayskip}{6pt}%
 }
 
\begin{document}

\title{
CO rotational line emission in very red carbon stars in the Magellanic Clouds
}  
 
\author{ 
M.~A.~T.~Groenewegen\inst{1} 
\and
A.~Nanni\inst{2,3} 
\and
M.~L.\ Boyer\inst{\ref{STSCI}}
\and
L.~Decin\inst{\ref{IvS}}
\and 
S.~R.~Goldman\inst{\ref{STSCI}}
\and
K.~Justtanont\inst{\ref{Chalmers}}
\and
F.~Kerschbaum\inst{\ref{Vienna}}
\and
I.~McDonald\inst{\ref{Jodrell}}
\and 
H.~Olofsson\inst{\ref{Chalmers}}
\and 
R.~Sahai\inst{\ref{JPL}}
\and
G.~C.~Sloan\inst{\ref{STSCI},\ref{UNorthC}} 
\and
J.~Th.~van~Loon\inst{\ref{Keele}}
\and 
W.~H.~T.~Vlemmings\inst{\ref{Chalmers}} 
\and
A.~A.~Zijlstra\inst{\ref{Jodrell}}
}

\institute{ 
Koninklijke Sterrenwacht van Belgi\"e, Ringlaan 3, B-1180 Brussels, Belgium \\ \email{martin.groenewegen@oma.be}
\and
National Centre for Nuclear Research, ul. Pasteura 7, 02-093 Warsaw, Poland
\and
INAF - Osservatorio astronomico d'Abruzzo, Via Maggini SNC, 64100, Teramo, Italy
\and
Space Telescope Science Institute, 3700 San Martin Drive, Baltimore, MD 21218, USA\label{STSCI}
\and
Institute of Astronomy, Department of Physics and Astronomy, University of Leuven, Celestijnenlaan 200D, 3001 Leuven, Belgium\label{IvS}
\and
Department of Physics and Astronomy, Chalmers University of Technology, SE-412 96 Gothenburg, Sweden\label{Chalmers}
  \and
  Department of Astrophysics, University of Vienna, T\"urkenschanzstr. 17, A-1180 Vienna, Austria\label{Vienna}
\and
Jodrell Bank Centre for Astrophysics,  The University of Manchester, Oxford Road, Manchester, M13 9PL, UK\label{Jodrell}  
\and
Jet Propulsion Laboratory, California Institute of Technology, MS\,183-900,  Pasadena, CA 91109, USA\label{JPL}
\and
  Department of Physics and Astronomy, University of North Carolina, Chapel Hill, NC 27599-3255, USA\label{UNorthC}
  \and
  Lennard-Jones Laboratories, Keele University, Staffordshire ST5 5BG, UK\label{Keele}
} 
 
\date{received: 2026, accepted: 2026} 
 
 
\authorrunning{Groenewegen et al.} 
\titlerunning{CO rotational line-emission in AGB stars in the LMC} 
 
\abstract
{
Stars of low and intermediate initial mass lose most of their stellar mass at the end of their lives during the asymptotic giant branch (AGB) phase.
Determining their gas and dust mass-loss rates (MLRs) is crucial for quantifying the contribution of evolved stars to the life cycle of dust and gas in the Universe.
The Atacama Large Millimeter/submillimeter Array was used to observe 38 carbon stars (C stars) in the large Magellanic cloud (LMC) and three C stars in the small
Magellanic cloud (SMC) in the CO J= 2-1 line.
     Line profiles were fitted to derive stellar velocities and wind-expansion velocities ($V_{\rm exp}$).
     CO emission is detected in two C stars in the SMC and 33 C stars in the LMC, extending earlier work from 2016 that detected CO emission from
       a total of four C stars observed in the LMC. This is the first detection of carbon monoxide around an AGB star in the SMC.
One object in the LMC shows emission in $^{13}$CO. The wind-expansion velocity ranges from  $\sim$7.5 to $\sim$30~\ks.

Archival data were used to determine the pulsation periods as well as construct and model the spectral energy distributions using two
dust radiative transfer codes.
      Mass-loss rates were independently derived from these two codes as well as from the intensity of the CO line, using a simple formula.
      On average, the dust-based MLRs higher  than the MLRs based on the CO line by a factor of 1.6.
      Additional CO data in other transitions, combined with proper modelling, is required to further investigate this possible discrepancy.
      Mass-loss rates, pulsation periods, and expansion velocities were compared to a sample of Galactic C stars. There is a strong bias,
      as the Magellanic Cloud targets    sample the highest MLRs and luminosities, yet they represent only a minority of stars in a Galactic sample.
      Comparing this sample with a similarly extreme set of Galactic stars with periods longer than 500 days, we identify no correlation between
      metallicity and either the MLR or $V_{\rm exp}$.
}

\keywords{Stars: AGB and post-AGB -- Stars: winds, outflows -- Radio lines: stars} 

\maketitle 
\nolinenumbers

\section{Introduction} 

Low- and intermediate-mass stars (with initial masses of $\sim$0.8--8~M$_{\odot}$) 
end their lives with an intense mass-loss episode during the asymptotic giant branch (AGB) phase. 
The standard picture is that stellar pulsation and dust formation drive a slow, cool wind.
This wind from AGB stars is one of the main sources that  enriches the interstellar medium (ISM) with gas and dust (see \citealt{Hofner18} for a review).
Carbon (C) stars are a subset of AGB stars that underwent so many helium shell flashes (or thermal pulses), a process in which carbon (the product of He burning)
mixes into the stellar envelope, that carbon atoms outnumber oxygen atoms (C/O $>$ 1). The initial mass range of AGB stars that become C stars depends on metallicity,
but is typically $\sim$1.5--4.0~M$_{\odot}$ (see e.g. \citealt{Ventura20,Ventura21}).

The dust mass-loss rate (MLR) is typically modelled using the spectral energy distribution (SED), which requires a (dust) expansion velocity, that cannot be
directly measured, and, therefore,  must be estimated. The gas MLR is usually determined from the rotational transitions
of carbon monoxide (CO), which includes a measurement of the expansion velocity obtained from the width of the line profile.
Wind-expansion velocities ($V_{\rm exp}$) have been measured for hundreds of AGB stars in the Milky Way (MW) through either CO thermal line emission, typically using
single-dish telescopes
(see representative studies by e.g. \citealt{KO99,Olofsson2002,Groenewegen2002a,Kemper03,DeBeck10,Ramstedt20,Wallstrom25}) or
OH maser line emission for O-rich sources (see the database by \citealt{Engels15})\footnote{\url{https://hsweb.hs.uni-hamburg.de/projects/maserdb_new/}}.

The first indication that the $V_{\rm exp}$ of AGB stars depends on environment (metallicity) was presented in \cite{Wood92} based on OH/IR stars in the
large Magellanic cloud (LMC).
For C stars this began with the detection of CO (2-1) emission in the halo object IRAS 12560+1656 \citep{Groenewegen97}, while 
\citet{Lagadec10} detected CO J= 3-2 emission in this and five other C stars. 
The $V_{\rm exp}$ were lower than those of C stars in the Galactic disk, with a value  as low as $\sim 3$ \ks\ in IRAS~12560.
Similar results have been derived for O-rich stars.  
\cite{McDonald2019} report a $V_{\rm exp}$ velocity of about 3.2~\ks\ in the metal-poor ([Fe/H]= $-0.72$) star 47 Tuc V3, while 
\cite{McDonald20} find a $V_{\rm exp}$ = 3.5~\ks\ in the metal-poor ([Fe/H] = $-1.3$ to $-1.0$ dex) star RU Vul.
These velocities are lower than that for typical O-rich AGB stars in the MW. Several factors may play a role, but one possibility is the lack of refractory
material at lower metallicity. However, since carbon stars produce their own refractory material through the dredge-up of carbon produced
during helium shell burning and thus therefore largely independent of the initial metallicity
it has been unclear whether their winds are similarly slowed.

Hydrodynamical wind models by \cite{Bladh19} predict MLR and $V_{\rm exp}$ for the small Magellanic cloud (SMC), LMC, and solar metallicities for a star with a current
mass of 1~\msol\ for a few effective temperatures, piston velocities, and carbon excesses.
Based on the  model outputs for 20 objects where data are available for all metallicities and for a piston velocity of 6~\ks, the MLR ratios are
solar/LMC= 0.87 $\pm$ 0.02 and solar/SMC= 0.83 $\pm$ 0.02 (mean value and error on the mean). The $V_{\rm exp}$ ratios are solar/LMC= 0.96 $\pm$ 0.02 and
solar/SMC= 0.96 $\pm$ 0.02. In other words, lower metallicities lead to slightly higher MLRs on average, exhibiting almost no dependence of $V_{\rm exp}$ on metallicity
(cf. Fig.~7 in \citealt{Bladh19}).
Gas-to-dust (GTD) ratios are not reported for the individual models but their Fig.~8 suggests values between 500 and 6000, independent
of metallicity.

Since maser emission is stronger than thermal emission, attempts have been made to detect OH emission in the 
SMC and LMC.
\cite{Goldman17} summarises the current state of affairs and presents accurate $V_{\rm exp}$'s 
for 13 OH/IR stars in the LMC. They suggest that $V_{\rm exp}$ is proportional to metallicity and luminosity as $L^{0.4}$.
However, \cite{Goldman18} still find no detections in the SMC.

\citet{Gr2016} (hereafter Gr16) presented the first detections of spectrally resolved CO line emission in extragalactic stars,
namely in the LMC. Four C stars were targeted
and all were detected, while two OH/IR stars were observed with one marginal detection.
A comparison was made to similarly highly red Galactic C stars, tentatively concluding that for C stars
the $V_{\rm exp}$'s in the LMC are lower than in the solar neighbourhood, while the MLRs appear similar. 

This paper presents observational results of a much larger sample of C stars both in the SMC and LMC to verify
some of the preliminary conclusions of Gr16.
Our selected list of targets is presented in the next section.
ALMA observations are presented in Sect.~3.
Selected observational results are presented in Sect.~\ref{S-ObsRes}, including an analysis of the CO line profiles.
The determination of the pulsation periods are discussed in Sect.~\ref{S-Puls}, while
the analysis of the SEDs and determination of the MLRs are presented in Sect.~\ref{Sec-Fitting}.
The paper ends with a discussion (Sect.~8) and a summary, with conclusions presented in Section~9.

\section{Targets}
\label{S-Tar}

Targets were selected based on predictions of the CO line strength, which, in turn, were based on detailed dust modelling and a comparison
with the results in Gr16.
Dust models are available for $\sim$8000 C-rich AGB stars in the LMC and $\sim$3000 in the SMC, based on mid-IR photometry from
the {\it Spitzer} SAGE survey (\citealt{Nanni18, Nanni19}, hereafter N18 and N19).
Models are also available for a subsample of $\sim$500 AGB and red supergiant (RSG) stars (including O-rich AGB stars), based on mid-infrared
spectra from the IRS on board {\it Spitzer} as well as from complete SEDs \citep{GS18}.

In the N18 and N19 models (also see Sect.~\ref{S-Nanni}), the growth of  dust grains of different chemical compositions in the circumstellar 
envelopes (CSEs) of thermally pulsing AGB (TP-AGB) stars is coupled with a  spherically symmetric, stationary wind (\citealt{Nanni2013, Nanni2014} and
references therein).
Applying this calculation to stellar evolutionary tracks reproduces the observed trend between the ($V_{\rm exp}$)
and MLR of Galactic C stars well \citep{Nanni2013}.
Dust radiative transfer (RT) calculations are performed a posteriori  to compute the spectra as reprocessed by the dust using the code
\textsc{More of Dusty} (MoD) \citep{Gr_MOD, Ivezic_D}.

This approach allows one to consistently compute the dust-production rate (DPR), the GTD ratio, and $V_{\rm exp}$ as well as the MLRs of the fitted stars.
The values obtained for the GTD ratio are comparable with results from detailed hydrodynamical calculations \citep{Mattsson10,Eriksson14}.

Based on the modelling results in N18 and N19, the CO-integrated intensities and peak line intensities 
were predicted using multi-parameter fits \citep{DeBeck10}, assuming parabolic CO profiles and the predicted  values for $V_{\rm exp}$.
Standard distances of 50 kpc for the LMC and 60 kpc for the SMC were assumed.
The predictions are relative intensities, as they also depend on the beam size and the CO/H$_2$ abundance.
These relative intensities were compared to the peak fluxes and integrated intensities for the four C stars detected in CO(2--1) by Gr16.
This comparison lead to a scaling factor, accurate to within a factor of 2, which enables us to select stars detectable with high certainty.
In addition, the formalism in \cite{DeBeck10} showed that the most efficient way to detect the CO line and determine the expansion velocity
is via  observations of the J=2--1 transition.  The 1--0 transition is weaker than 2--1. The 3--2 line is slightly stronger; nonetheless, it is not 
strong enough to compensate for the less favourable atmospheric transmission.

The initial sample was restricted to C stars with a predicted CO(2--1) intensity greater than 20 mJy (i.e. half the intensity
of IRAS~05506$-$7053 in Gr16). This sample was then further refines to cover a range of predicted $V_{\rm exp}$, GTD ratios, $L$,  and MLR,
as well as to ensure the availability of IRS spectra and pulsation periods. 

The final sample consists of three SMC and 38 LMC C stars. Combined with the four C stars in Gr16, this represents only 0.5 percent
of the  total estimated population of C stars in the LMC. Nevertheless, the combined MLRs of these stars is estimated
to be 8 $\times 10^{-4}$ \msolyr\ (or 2.3 $\times 10^{-6}$ \msolyr\ in dust), or $\sim$ 6\% of the total gas and 16\% of the total dust
returned by the entire population \citep{Nanni19}. This indicates  the importance of studying in detail this representative sample of the reddest stars.

To quantify the redness of our sample, we note that \citet{Boyer2011} in their study of evolved stars in the LMC observed with {\it Spitzer}
found 25914 objects with detections in the [3.6] and [4.5] bands (excluding red giant branch (RGB) and RSGs).
The LMC star in our sample with the bluest colour has [3.6-4.5] = 0.37; the second bluest has 0.61, and the reddest colour is 3.3~mag.
In the non-RGB and non-RSG samples of  \citet{Boyer2011} 1016 objects have  [3.6-4.5] $>0.37$~mag, corresponding to approximately 4\%.
Our stars are chosen to be carbon stars, so restricting the  \citet{Boyer2011} sample to the relevant classes of C-AGB, x-AGB and FIR, this ratio
becomes 992 out of 8761 objects. Our sample of very red C stars, therefore, represents about the reddest 10\% of the known and likely C stars in the LMC.

\begin{table*}
\setlength{\tabcolsep}{1.0mm}

 \caption{ALMA SGs and general properties.}
\footnotesize
  \begin{tabular}{lcrcccrc}
  \hline\hline
SG           &    Observation dates                & EB & T      & N$_{\rm Tel}$   &  Baselines  & N$_{\rm obj}$   &           Beam              \\
             &                                     &    & (h:m)  &                &   (m)       &               & (\arcsec\ $\times$ \arcsec)   \\
\hline
SMC          & 2024-01-01 21:36 - 2024-01-02 00:49 &  3 & 1:49   & 46             & 15 - 784                    &  3 & 0.585 $\times$ 0.496 \\
LMC strong   & 2024-01-02 05:14 - 2024-01-02 01:55 &  2 & 1:75   & 46,49          & 15 - 784                    & 16 & 0.570 $\times$ 0.516 \\
LMC weak     & 2024-01-04 03:26 - 2024-05-24 18:55 & 11 & 8:52   & 41,43,44,46,47 & 15 - (284,312,500,740,784)  & 22 & 0.826 $\times$ 0.654 \\
\hline
\end{tabular}
  \tablefoot{Listed are the name of the science goal, range in observation dates, number of execution blocks, total time on target,
    the number of telescopes used, the range in baselines, the number of science targets, and the typical size of the reconstructed beams.} \\
\label{Tab-SG}
\end{table*}

\section{Observations and post-processing of the data} 
\label{S-Obs}

The predicted CO strengths cover a range in intensities, and we therefore decided to split the observations into three science goals (SGs):
one for the SMC stars, one for the LMC stars predicted to have weaker emission (called LMC weak or LMCwk), and one for the LMC stars
predicted to have stronger emission (called LMC strong or LMCst).
For the first two SGs an rms noise level of 3~mJy was requested. For the third SG, the requested noise level was 6~mJy.

The sources were observed with ALMA at Band 6 in early 2024. Table~\ref{Tab-SG} provides the details of the observations
as well as the typical size of the reconstructed beams.
The average on-source integration time was 7 minutes for LMCst  and 24 minutes for the other two SGs.

The observations had four spectral windows (SPWs): 
one with a width of $937.5$~MHz and 1920 channels to cover the $^{12}$CO(2--1) transition centred on 230.538~GHz, 
one window with the same spectral set-up to cover the    \mbox{$^{13}$CO(2--1)} transition centred on 220.399~GHz, 
and two $1.875$~GHz windows with 240 channels each for the continuum centred on 219.00 and 232.80~GHz, respectively.
The channel spacing in the line SPWs was about 0.64~km~s$^{-1}$ corresponding to an effective spectral resolution of about 0.74~\ks.
The image cell size was 0.120\arcsec\ for the LMCwk SG, and 0.096\arcsec\ for the two  other SGs.

Pipeline products were retrieved from the ALMA archive that had been processed using the Common Astronomy Software Applications
(CASA, version 6.5.4.9) pipeline \citep{McMullin07}.
The quality control log files indicated that self calibration was attempted but not applied due to the low signal-to-noise.
Post-processing was done on local computers using the same version of CASA.

The first step in the post-processing aimed at determining the source position and shape of the source based on the $^{12}$CO data.
A zero-moment map was created using channels 800-1200 (covering $\sim$95 to $\sim$378~\ks) for the LMC sources and
channels 700-1100 (covering $\sim$50 to $\sim$305~\ks) for the SMC sources, corresponding to the expected range in velocities  showing emission.
A rectangular region of 41~pixels wide (about 4-5\arcsec$\times$4-5\arcsec) placed in the centre of the moment map was defined and a two-dimensional
Gaussian fit was performed resulting in the position of the source (with error bars) and the major and minor axis and position angle
(with error bars) of the Gaussian.
It was verified a-posteriori that the input coordinates based on SAGE IR data matched very well with the ALMA position, with a difference of
at most 0.5\arcsec, and a median difference of 0.20\arcsec\ among 35 detections, so that the sources are indeed expected to be located in the
centre of the image.

The next step was to extract the $^{12}$CO line spectrum.
The size of the region, from which the spectrum was extracted, was set by the (maximum) expected size of the CO emitting region.
The photo dissociation radius $r_{\frac{1}{2}}$, defined as the radius where half the CO is dissociated,
is larger for higher MLRs, lower $V_{\rm exp}$, and larger $f_{\rm CO/H_2}$ values.
The largest value in the grid of \citet{Saberi19} is about 1.1 $\times$ 10$^{18}$~cm for an expansion velocity of 7.5~\ks,
\mdot= 1 $\times$ 10$^{-4}$~\msolyr\, and  $f_{\rm CO/H_2}$= 10 $\times$ 10$^{-4}$.
At 50~kpc this implies a (maximum) CO shell size of about 1.4\arcsec.
For an expansion velocity of 15~\ks, \mdot= 2 $\times$ 10$^{-5}$~\msolyr\, and  $f_{\rm{CO/H_2}}$= 8 $\times$ 10$^{-4}$, it
is already 3.5 times smaller.
About half of the sources have predicted MLRs in the range 2-4 $\times$ 10$^{-5}$~\msolyr, based on N18 and N19.
%
As we discuss deriving MLR estimates later in the analysis, one could fine-tune this aperture on a source-by-source basis; however,
this would introduce large uncertainties. Therefore we chose to adopt a conservative large aperture size.
Given a spatial full-width half-maximum (FWHM) major axis of the beam of (at most) 0.83\arcsec\ and a maximum photo dissociation radius of 1.4\arcsec, 
the line flux was extracted inside a circle with radius 2.0\arcsec.
This aperture should cover the line-emitting region in all the objects and is centred on the pixel coordinates of the object.
A Gaussian fit was made to the profile to obtain a peak velocity estimate and the FWHM width of the profile.

In a third step, the $^{13}$CO(2--1) profile was extracted. Anticipating that all but one source would be non-detections, the velocity range
of the $^{12}$CO profile was converted to the range of channels in the $^{13}$CO SPW, and the line was extracted within the aforementioned spatial  region.

The fourth step involved determining the dust-continuum emission in the line-free SPWs.
The four SPWs were treated separately.
For each SPW, a $-1$st moment map was created (i.e. giving the average value along the spectral axis), excluding the channels with emission.
Although only one source was clearly detected in $^{13}$CO, the expected spectral region for the $^{13}$CO emission was excluded in all stars based
on the spectral range of the $^{12}$CO detection.
The region where the dust emission can occur is larger than the CO-emitting region.
Based on the RT calculations mentioned in Sect.~\ref{S-Tar}, the outer radius of the dust shell, where the dust temperature becomes 20~K
is about 4.5\arcsec\ at the distance of the LMC for the objects with the largest predicted MLRs.
The continuum emission was extracted from a circular region of radius $r_{\rm source}$= 4.5\arcsec\ centred on the source (about 38~pixels).
As for the aperture to derive the CO emission, one might fine-tune the aperture to estimate the continuum emission on a source-by-source basis. 
We, nevertheless, chose to adopt a conservative large aperture size for all sources.

An off-source region was defined as a circle with an inner and outer radius that are (arbitrarily) 4 and 24~pixels larger than $r_{\rm source}$.
From that region, the rms noise was extracted (the median absolute deviation times 1.483\footnote{This gives the 1$\sigma$ value in a Gaussian distribution.} was
  actually used as a more robust estimate of the noise level), and the error in the flux in the on-source region was determined.

This procedure produces four estimates of the continuum flux with error bars at four wavelengths, typically ranging between 1290 and 1370~$\mu$m.
From the RT calculation, the flux  from C stars at millimetre wavelengths is expected to follow $\lambda^{-2}$; therefore, the four flux estimates
were scaled to a wavelength of 1330~$\mu$m. Then, the weighted average and the error in the mean were calculated.
If the flux was less than three times the error bar, the 3$\sigma$ upper limit was quoted.

Finally, a 10\arcsec$\times$10\arcsec\ cut-out of the $^{12}$CO region was produced, based on the zeroth moment image (see Appendix~\ref{App-Mom0}).
This step was made in a second iteration, where the fixed range in the channels was replaced with a range in velocity centred on the detected line (when detected).

\section{Observational results} 
\label{S-ObsRes}

\subsection{Analysis of the line profiles}
\label{SS-Prof}

The line profiles were fitted with a `Shell' profile as defined in the CLASS/GILDAS software
package\footnote{\url{https://www.iram.fr/IRAMFR/GILDAS/doc/html/class-html/node38.html}}:

\begin{equation}
   P(V) = \frac{A}{\Delta V \; (1 + H /3)} \; \left(1 + 4 H \; \left(\frac{V - V_{\star}}{\Delta V}\right)^2\right),
\end{equation}
where $V_{\star}$ is the stellar velocity (in \ks; throughout the paper the Local Standard of Rest (LSR) frame is used),
$A$ is the integrated intensity (in \jks), 
$\Delta V$ the full-width at zero intensity (in \ks; $V_{\rm exp}$ is taken as half that value), and
$H$ is the horn-to-centre parameter.
This parameter describes the shape of the profile. It is $-1$ for a parabolic profile, $0$ for a flat-topped one, and $>0$ for a double-peaked profile.
$H$ was fixed to $-1$ in the fitting.
A parabolic profile indicates optically thick, unresolved emission.
Leaving $H$ as a free parameter typically gives a value with a large error bar consistent with $-1$.
Table~\ref{Tab-final} lists the results and all profiles are shown in Appendix~\ref{App-LP}.

Error bars on $V_{\star}$ and $\Delta V$ are based on Monte Carlo simulations, in which simulated spectra were generated by using
a Gaussian distribution of the errors and subsequently re-fitting the profile. The quoted parameters represent the median value, with
the error bars given by the difference between the 69\% and 31\% percentiles. Upper limits are based on the 97.3\% percentile (i.e. 3$\sigma$ values).
In the case of the non-detected $^{13}$CO lines, the expansion velocity was fixed to the value derived from the $^{12}$CO profile, and
the systemic velocity was allowed to vary within the width of the $^{12}$CO profile.
In the case of $^{12}$CO non-detections, the upper limit to the intensity was derived by fixing $V_{\rm exp}$ to 9~\ks\ for the SMC and
15~\ks\ for the LMC sources (the typical values among the detections) and allowing
the stellar velocity to vary relative to an arbitrary initial value.

\subsection{Non-detections}

Thirty-five out of forty-one objects were detected.
For one of the three SMC and five of the 38 LMC sources, no convincing $^{12}$CO profile could be derived.
For four objects (SAGE MCJ045344, MSX LMC 1220, MSX LMC 474, and MSX LMC 527) the images in Appendix~\ref{App-Mom0} appear to
show a possible source close to the centre; however, the spectra (Figure~\ref{App-Fig-LP}) revealed no line,
even after they were binned in velocity space.

\begin{figure}
  \centering
\begin{minipage}{0.45\textwidth}
\resizebox{\hsize}{!}{\includegraphics[angle=0]{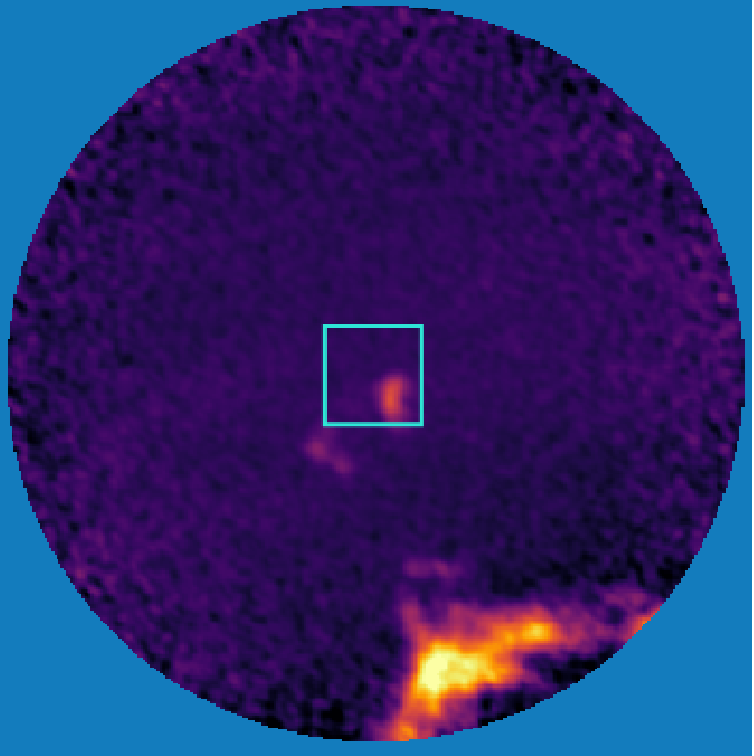}} 
\end{minipage}

\caption{\label{Fig-527} Emission in five spectral channels in the field around MSX LMC 527 (about 36\arcsec\ across).
The box in the centre measures about 5\arcsec\ per side and is centred on the nominal source position.}
\end{figure}

The observation of MSX LMC 527 is contaminated by emission from the ISM. Figure~\ref{Fig-527} shows an hourglass
structure unrelated to the object. The emission is extremely narrow (five channels or about 3~\ks) and
much stronger than any of the other detections. The  CO profile shown in Figure~\ref{App-Fig-LP} excludes these channels, and no obvious
underlying emission related to the AGB star is visible. Other ISM emission appear in the field at these velocities, which also indicates
that this emission is not centred on the expected source position.

Interstellar medium contamination also occurs in the detected source MSX LMC 1780, where seven channels have been removed in the profile
shown in Figure~\ref{App-Fig-LP}. The removal of these channels did not impact the derivation of the parameters of the profile.
In the cases of SAGE MCJ045344, MSX LMC 1220, and MSX LMC 474, ISM contamination does not seem to play a role.
Observations with better sensitivity, or the use of a prior on stellar velocity (e.g. from infrared spectroscopy), could reveal the CO profile
of the AGB star or yield a more stringent upper limit.

\subsection{The $^{13}$CO detection}

One of the targets is detected in $^{13}$CO, the LMC carbon star IRAS 05568 (see Fig.~\ref{App-Fig-LP}).
The ratio of the integrated intensities is $^{12}$CO/$^{13}$CO= 1.8 $\pm$ 0.1. The $^{12}$CO is likely optically thick, as indicated by the parabolic line shape;
therefore, the $^{12}$C/$^{13}$C ratio is likely larger but probably not by a huge factor. To quantify this, full RT modelling would be required.
The $^{12}$C/$^{13}$C ratio is similar to that observed in J-type C stars (4-10,  \citealt{AbiaIsern2000} close to the CN cycle
equilibrium value of 3). J-type C stars are not rare in the MW and LMC (10-15\%;  \citealt{AbiaIsern2000, Morgan2003}) but the J-type
character or quantitative abundance analysis has been done exclusively based on optical spectra so far.
This observation is the first indication of the J-type nature of an LMC C star with a high MLR, and the second object after the
galactic object IRAS 15194$-$5115 ($^{12}$CO/$^{13}$CO= 6, \citealt{Woods03}).
Among the $^{13}$CO non-detections, the most stringent lower limits come from stars where the $^{12}$CO is strong,
$^{12}$CO/$^{13}$CO $\gtrsim$ 15-19 (IRAS 05416, TRM 74, and IRAS 05190).

\section{Pulsation periods}
\label{S-Puls}

Time-series data for the objects were obtained from the VISTA Magellanic Cloud (VMC) survey \citep{Cioni11} Data Release (DR) 6
in the $K$ band as well as the Wide-field Infrared Survey Explorer (WISE) and Near-Earth Object WISE (NEOWISE) surveys
\citep{Wright10,Mainzer11,Mainzer14} in the W1 and W2 bands (centred at 3.4 and 4.6~$\mu$m, respectively).

From the AllWISE multi-epoch photometry table and the NEOWISE-R single-exposure source
table\footnote{see \url{https://irsa.ipac.caltech.edu/cgi-bin/Gator/nph-scan?mission=irsa&submit=Select&projshort=WISE}}, all entries
within 1\arcsec\ of the source coordinates were downloaded in the W1 and W2 filters, with the additional constraint that the error bars on the magnitudes
were $<0.25$~mag and the flags {\it saa\_sep} $>5$ and {\it moon\_masked} $=0$ (e.g. \citealt{Uchiyama19}) were applied.

At the bright end, the WISE and NEOWISE data suffer from saturation that influences the photometry; thus, a correction was applied.
Table~2 in Sect.~II.1.c.iv.a of the NEOWISE Explanatory Supplement\footnote{\url{http://wise2.ipac.caltech.edu/docs/release/neowise/expsup/sec2_1civa.html}}
contains correction tables in the W1 and W2 filters for all phases of the mission.
At the bright end (fourth mag) the corrections are about 0.06 and 0.4~mag in the WISE W1 and W2 filters and 0.7 and 0.9~mag in the NEOWISE W1 and W2 filters.
These corrections reach low (0.02~mag) levels at 6.8 and 6.0~mag (WISE W1 and W2) as well as 6.9 and 7.9~mag  (NEOWISE W1 and W2), respectively.
The light curves (LCs) were pre-analysed with the code {\sc Period04} \citep{Period04} to estimate the period.
With this initial guess for the period, a function, 
\begin{equation}
\label{Eq-fit}
m(t) = m_0 +
A \sin (2 \pi \; t \; e^{f}) +
B \cos (2 \pi \; t \; e^{f}), 
\end{equation}
was fitted to the LC using the weighted linear least-squares fitting routine {\sc mrqmin}, assuming  a single dominant period for simplicity.
Fitting was performed with the Fortran codes available in {\it Numerical Recipes} \citep{Press1992}, as
described in Appendix~A of \citet{Groenewegen04}, and modified to analyse VMC $K$-band data  and
WISE/NEOWISE data, as described in \citet{Groen2020,Groen2022}.
The resulting parameters are listed in Tables~\ref{Tab-VMC-Periods} and ~\ref{Tab-WISE-Periods} for the VMC and WISE data, respectively.
These include the mean magnitudes ($m_0$), the period ($\exp (-f)$), and amplitude ($\sqrt{A^2 + B^2}$) with their associated uncertainties.
Figures~\ref{AppFig-VMC} and \ref{AppFig-WISE} provide examples of the LCs and the fits, and the full set is available through Zenodo.
The periods derived in the present paper supersede those reported in Gr16 for IRAS 05506, IRAS 05125, ERO 0529379, and ERO 0518117, as they are based
on additional data.
Table~\ref{Tab-SED} lists the adopted period in the last column. This is determined as the weighted mean (and the error on the mean) of the available periods.
The column is empty when the period is larger than 2500~d, the amplitude is smaller than 0.2~mag, or when the error on the amplitude is greater
than $\frac{1}{3}$ of its value.

\section{SED fitting and mass-loss rates}
\label{Sec-Fitting}

Two sets of RT dust models have been used, as described in the following sections.
Both have complementary approaches, in order to better appraise uncertainties in the derived parameters.

\subsection{More of DUSTY models}
More of Dusty models were run using the `density type = 3' mode, in which the hydrodynamical equations of dust and gas are solved and
the gas-expansion velocities and gas MLRs are predicted (following Gr16).

Inputs to the RT model include atmosphere models from \citet{Aringer09} and a combination of amorphous carbon (AMC), silicon carbide (SiC),
and magnesium sulphide (MgS) dust. Details are given in  Appendix~\ref{App-SED}.

The RT models were fitted to broad-band photometry (compiled in Table~\ref{Tab-Phot}) and {\it Spitzer} IRS spectra (see \citealt{GS18} for details), when available.
The fitted parameters include luminosity, dust optical depth ($\tau_{\rm d}$ at 0.5~$\mu$m), and the dust temperature at the inner radius, $T_{\rm c}$. 
The outer radius was set to 3000 times the inner radius.
The velocities and MLR, given by {\sc dusty}, scale as
\begin{equation*}
  V_{\rm exp} \sim (L/10^{4})^{0.25} \left((r_{\rm gd}/200)(\rho_{\rm d}/3)\right)^{-0.5}  {\rm and}
\end{equation*}
\begin{equation*}
  \dot{M} \sim (L/10^{4})^{0.75} \left((r_{\rm gd}/200)(\rho_{\rm d}/3)\right)^{+0.5} , {\rm respectively,}
\end{equation*} 
where $L$ is the luminosity in solar units, $r_{\rm gd}$ is the GTD ratio, and $\rho_{\rm d}$ is the specific density of 
the dust grains in gram~cm$^{-3}$.

The GTD ratio was first tuned to fit the observed expansion velocity (taking the specific density for the adopted dust species of that source), and then
used with the fitted luminosity to derive the MLR.
The error on the GTD ratio was calculated accounting for the error in $V_{\rm exp}$ and $L$. The error in the MLR accounts for
the error in the GTD ratio, the error in the dust optical depth, and the error in luminosity.
Additional uncertainties are hard to account for, which relate to differences in grain optical properties and grain densities. 
Fitting results are listed in columns~7--10 in Table~\ref{Tab-SED}.
Examples fits to the SEDs are  discussed in Sect.~\ref{S-ALMA} and shown in Fig.~\ref{AppFig-SED}.

\subsection{Nanni models}
\label{S-Nanni}

Second, for the SED fitting, we adopted the grids of models presented in N18 and N19 initially used to predict the intensities described in Sect.~\ref{S-Tar}.
In these models, dust formation is self-consistently computed throughout the CSE of the star, coupled with stationary wind, given the stellar input parameters 
(e.g. current stellar mass, effective temperature, luminosity, MLR, and photosphere element abundances), as fully described
in \citet{Nanni2013, Nanni2014} (see Appendix~\ref{App-SED}).

The models predict $V_{\rm exp}$, the GTD ratio, the dust temperature at the inner boundary of the CSE as well as the dust density profile. 
These outputs of the calculation were adopted as inputs to the MoD code to compute the spectra reprocessed by dust as a function of the stellar parameters.
The adopted model parameters were selected to have predicted $V_{\rm exp}$'s within 3$\sigma$ of the observed values, while all the input parameters
were free to vary.
The fitting results are listed in columns~2--6  of Table~\ref{Tab-SED}.

A comparison between the MoD and the Nanni models is presented in Appendix~\ref{App-SED}.
The geometric mean of the independent model calculations was adopted, and the values are given in columns~11--13 in Table~\ref{Tab-SED}.

\begin{table*}

 \caption{Highly red Galactic carbon stars. }
  \centering
  \begin{tabular}{lrcccccrcc}
  \hline \hline
Identifier  &   Period  &  $V_{\rm exp}$ &   Lum.  &   D   & MLR$_{\rm DUSTY}$   &  $V_{\rm exp, DUSTY}$ & GTD        &  MLR$_{\rm scaled}$     \\
            &   (days)  &    (\ks)     & (\lsol) & (kpc) & (10$^{-5}$\msolyr) &       (\ks)       &            &  (10$^{-5}$\msolyr)     \\
\hline
AFGL 190        & 1060 &  18.0 & 16~400 & 3.30 & 5.7 & 10.5 &  67 & 3.3 \\ 
AFGL 341        &  815 &  14.2 & 12~500 & 2.81 & 2.6 & 14.2 & 199 & 2.6 \\ 
IRAS 03448+4432 &  729 &  13.3 & 11~100 & 2.31 & 1.4 & 20.6 & 477 & 2.2 \\ 
AFGL 865        &  696 &  16.6 & 10~600 & 1.72 & 1.5 & 17.9 & 234 & 1.6 \\ 
IRAS 08074-3615 &  832 &  21.7 & 12~700 & 2.95 & 2.0 & 19.2 & 156 & 1.7 \\ 
AFGL 2494       &  783 &  20.5 & 12~000 & 1.50 & 1.4 & 20.0 & 191 & 1.3 \\ 
AFGL 3068       &  696 &  15.1 & 10~600 & 1.05 & 3.9 & 13.9 & 168 & 3.6 \\ 

\hline
\end{tabular}
\tablefoot{
Data from Table~4 in Gr16 where references are given.
Columns~6 and 7 give the gas MLR and gas expansion velocity for a GTD ratio of 200 as outputted by DUSTY.
Column~8 gives the GTD ratio that makes the modelled expansion velocity equal to the expansion velocity, and the last column lists 
the gas MLR for that GTD ratio.
}
\label{TabExtr}
\end{table*}

\subsection{SED fit results and ALMA continuum data points}
\label{S-ALMA}

Examples fits to the SEDs from the two models are shown in Fig.~\ref{AppFig-SED}.
The comparison can be considered acceptable in some cased, such as for MSX LMC 527 and IRAS 06018 in Fig.~\ref{AppFig-SED}, given that the two models
have fundamentally different approaches (see discussion in Appendix~\ref{App-SED}).
However, it is poor in other cases (e.g. IRAS 05315).

It is illustrative to discuss the cases comparing the observed versus predicted ALMA 1330~$\mu$m continuum.
For 32 objects an upper limit was derived, and in 27 cases the MoD model indeed falls below the upper limit, such as  for MSX LMC 527 and IRAS 06018.
In three cases the discrepancy is deemed acceptable (within a factor of 2; e.g. IRAS 05568), but for two cases the difference is
large (TRM 74 and IRAS 05315).
Among the nine detections, the model agrees very well with the observations for one source (IRAS 05133) and predicts somewhat lower fluxes for six sources.
However, in one case (IRAS 05495) the model significantly overestimates the flux.

\begin{figure*}[h]
  \centering

  \includegraphics[width=0.32\hsize,angle=-0]{IRAS05315_sed.ps} 
  \includegraphics[width=0.33\hsize,angle=-0]{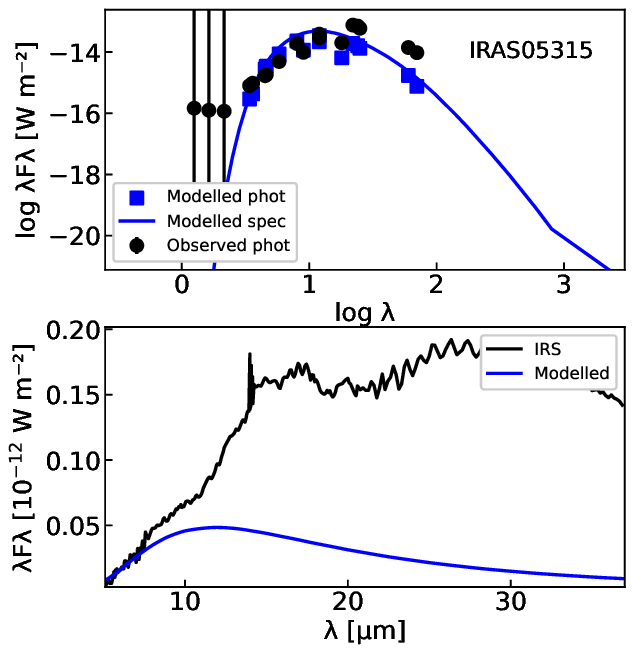} 
  \includegraphics[width=0.32\hsize,angle=-0]{IRAS05315_sed_F.ps} 

  \includegraphics[width=0.32\hsize,angle=-0]{IRAS05495_sed.ps}
  \includegraphics[width=0.33\hsize,angle=-0]{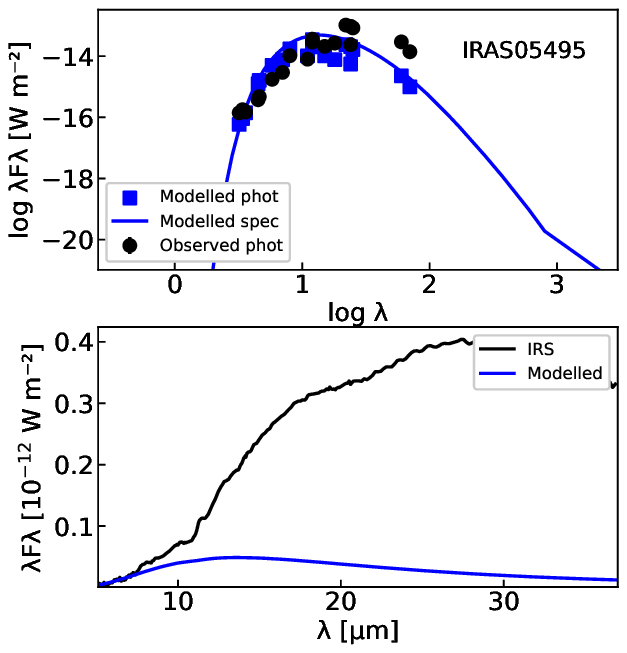} 
  \includegraphics[width=0.32\hsize,angle=-0]{IRAS05495_sed_F.ps} 

  \caption{\label{Fig-SED} Standard SED fits from the MoD (left) and Nanni (middle panel) models, and a MoD model with an adopted density law ($r^{-2.5}$)
    tuned to ﬁt the ALMA continuum at 1330 µm (right panel) for IRAS 05315 (top row) and IRAS 05495 (bottom row). See Sect.~\ref{S-ALMA} for details.
 }
\end{figure*}

IRAS 05315 and IRAS 05495 are good illustrations of how different assumptions in the MoD and Nanni models lead to very different results.
They likewise illustrate that the standard MoD model fails to fit the ALMA dust continuum observation (see Fig.~\ref{Fig-SED}).
In the Nanni models, the dust temperature at the inner radius reflects the dust condensation process, and the temperature is of the order of
950~K for these two objects. In the MoD models, this temperature is a free parameter and is found to be low (around 300~K).
In the Nanni models, the best fit under-predicts the mid-IR flux,
while the MoD model provides a slightly better overall fit but over-predicts the mid-IR and millimetre fluxes.

One way of lowering the millimetre flux is to have less dust than that predicted by the radiation-driven wind model where the dust density follows 
$r^{-2}$ from the point where the terminal wind velocity is reached. The right-hand panels of Fig.~\ref{Fig-SED} present models with a $r^{-2.5}$ dependence.
They clearly yield much-improved fits but result in current MLRs (i.e. at the dust inner radius) that are significantly higher
(of order $3 \cdot 10^{-4}$ \msolyr) than those in the $r^{-2}$ case ($\sim 2 \cdot 10^{-5}$ \msolyr).
Tuning the outer dust radius to where the dust reaches 20~K implies an increase in the MLR by factors of $\sim$15 over the past 22-34 kyr for these two stars.

One interpretation of the lower temperature at the inner dust radius is that the MLR recently stopped and that the CSE now drifts outwards.
Comparing the inner radius to that with dust temperatures of $\sim$950~K at the inner radius implies a flow timescale of 110-150~yrs for these two stars.
This is short compared to the typical post-AGB transition timescale of a few thousand years \citep{Bertolami16}; thus, having 
three objects (TRM 74 as well) in our sample would be statistically unlikely.

Another possibility to explain the very poor fits is a deviation from spherical symmetry of the CSE and/or the role of binarity.
Disk and spiral structures can manifest themselves in the SED, as shown  by \citet{Wiegert20} in one particular case.
That (sub)stellar companions can shape the molecular outflows is well established for MW AGB stars (see the review by \citealt{Decin20}), but the
current observations lack spatial resolution to study this.
Companions can also change the shape of the low-$J$ CO lines \citep{Vermeulen25}; however, at the current signal-to-noise ratio as well as the
given spectral and spatial resolutions,
the profiles show the standard parabolic profile (see Fig.~\ref{App-Fig-LP} for IRAS 05315, and the supplementary material for IRAS 05495 and TRM 74).

\subsection{MLR from simple CO scaling relation }

The relation in \citet{Ramstedt08} was used to determine the total MLR from the observed CO intensity: 
\begin{equation}
\dot{M} = s_{\rm J} \left( A \, \theta^2 \, D^2 \right)^{a_{\rm J}} {\rm V}_{\rm exp}^{b_{\rm J}} \, f_{\rm CO}^{c_{\rm J}}
\label{Eq-CO}
\end{equation}
with $s_{\rm J}= (1.3 \pm 0.7) \cdot 10^{-11}$~\msolyr, $a_{\rm J}= 0.82$, $b_{\rm J}= 0.46$, and $c_{\rm J} = -0.59$ for the J= 2--1 transition.
The distance, $D$, was set to 50~kpc for the LMC and 60~kpc for the SMC, and the averaged beam size ($\theta$) was taken as the geometric average
of the values in Table~\ref{Tab-final}.
The integrated intensity, $A$, in this formula is expressed units of Kelvin~\ks. The values in Table~\ref{Tab-final} are in Jansky~\ks\ and are converted.

The unknown in the equation is the abundance ratio of CO relative to H$_2$.
In Gr16 a value of \fco of $4.5 \cdot 10^{-4}$ was used for the LMC.
New calculations (Marigo \& Groenewegen, unpublished) confirm that, when the  object becomes a C star, the CO abundance is nearly constant
over the remainder of the AGB evolution. For a star with an initial mass of 2~\msol, this value is 10.8, 6.0, and 3.0 $\cdot 10^{-4}$
for initial abundances of $Z$= 0.014, 0.008, and 0.004, respectively. The change over the C-star's lifetime is only about 0.1 $\cdot 10^{-4}$.
Therefore, 3.0 and 6.0 $\cdot 10^{-4}$ were adopted for $f_{\rm CO}$ in the SMC and LMC, respectively.
The resulting MLR estimates are listed in the column~14 of Table~\ref{Tab-SED}.

Errors in the MLR estimate were calculated based on the error in the leading coefficient, $s_{\rm J}$, the error in the observed intensity, the error
in the expansion velocity, and an error of $1 \cdot 10^{-4}$ in $f_{\rm CO}$.    The error in $s_{\rm J}$ dominates, and the total relative error
is quite uniform within 54 and 59\%. The error in the CO-based MLR is large, but this is inherent when using only a single CO line and a
simple fitting formula.

\begin{figure}

  \centering
  
\begin{minipage}{0.45\textwidth}
\resizebox{\hsize}{!}{\includegraphics[angle=-0]{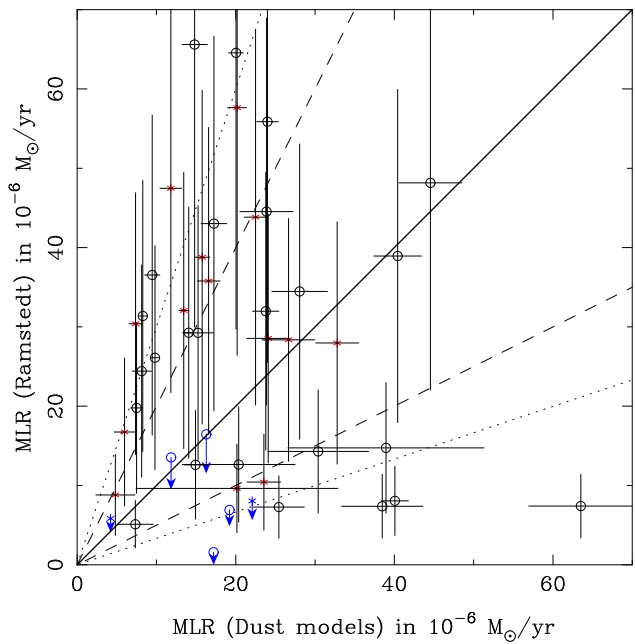}} 
\end{minipage}

\caption{\label{Fig-RTCO}  Comparison of the MLRs from dust RT ﬁtting and CO (2-1) line emission.
  The solid line shows the one-to-one correspondence between these MLRs, and the dashed (dotted) lines indicate factors of 2 (3) diﬀerence
  in the MLRs.   Upper limits on the gas MLR derived from CO non-detections are marked in blue.
  The red stars indicate unreliable models (see details in Appendix~\ref{App-SED}).
}
\end{figure}

Figure~\ref{Fig-RTCO} compares dust-based and CO-based MLRs. The errors in the CO-based MLR dominate the plot, and there is  very large scatter.
A straight average  of the 26 objects with CO detections and reliable RT models indicates that the dust-based MLRs
are, on average, a factor 1.6  larger than the CO-based ones, and a factor of 0.8 in the median.
The difference between the mean and the median also indicates  large scatter.
Additional CO data in other transitions combined with proper modelling of the CO lines is required to further investigate this possible discrepancy.

\section{Discussion} 

Figure~\ref{Fig-Comp} compares the results for the LMC and SMC to observations of Galactic objects (cf. Figure~2 in Gr16).
The Galactic sample was compiled from \citet{SO2001}, \citet{RO14}, and \citet{Danilovich15}.
As in Gr16, seven of the reddest known Galactic C stars, partly with SiC in absorption, were added (see Table~\ref{TabExtr}; data
were taken from Table~4 in Gr16). 
Apart from redness, these stars were selected to have a pulsation period in order to derive a luminosity from the $PL$-relation
in \citet{GW96} so that a distance could be estimated.
This unfiltered comparison reveals very little overlap between the Galactic and MC sample in most of the parameters.
It is clear that the MC samples longer periods (and higher luminosities).
Figure~\ref{Fig-CompP} shows the same plots but where the Galactic sample was limited to stars with a period longer than 500~days ($\log P > 2.7$).
This is about the shortest period in the MC sample (see Table~\ref{Tab-SED}).
The overlap in properties has improved. From the $V_{\rm exp}$ -- $\log P$ diagram we find no difference between the LMC and MW, although the two
SMC sources appear to exhibit a lower expansion velocity at that period. Comparing this and Fig.~\ref{Fig-Comp}, it is striking to see that, in the LMC,
our CO observations sample a population of intermediate luminosities ($\log L \sim 3.6-3.85$, $L \sim 4000-7000$~\lsol) with large MLRs that have no counterpart
in the Galactic sample considered here.
As a C star in the LMC has greater carbon excess than a MW C star at the same luminosity (due to lower oxygen to overcome in the LMC) its MLR may well be higher.

The comparison of MW and MC samples remains problematic, and this complicates quantifying with certainty any metallicity dependence of MLRs, GTD ratios, and $V_{\rm exp}$.
The Nearby Evolved Stars Survey (NESS; \citealt{SciclunaNESSii}) aims to construct a volume-complete sample of MW AGB stars.
Their samples are divided into several tiers based on the DPR and out to a limiting distance, which differs per tier.
AFGL 865, 2494, and 3068 are present in the latest version of the catalogue \citep{McDonald25}\footnote{The file table-g6.tsv
from the supplementary-tables directory as retrieved from the MNRAS website.}, specifically in the  'high' tier.
This tier indicates $3 \cdot 10^{-9} <$ DPR $<1 \cdot 10^{-7}$ \msolyr and a limiting distance of 1.2~kpc. Indeed, these three stars are closest
to the Sun according to our distance estimates, although two lie beyond 1.2~kpc.
The distances in \cite{McDonald25} have been updated for individual objects compared to \cite{SciclunaNESSii},
and the stars in the 'high' and 'extreme' tiers now extend to several kiloparsecs.
In addition, our DPR for AFGL 3068 would place it in the 'extreme' tier.
Considering this, the four stars in Table~\ref{TabExtr} should also have been included in NESS, but they were not.
This simply illustrates how difficult it is to create volume-complete samples given the uncertainties in distances (and, thus, luminosities and MLRs)
for MW AGB stars.
An additional complication in comparing samples comes from the fact that stars with high DPRs are rare.
Using the number of stars per tier from \cite{McDonald25} and excluding 5\% of stars with the largest outlying distance per tier,
assuming a volume density proportional to the square of
the distance\footnote{As the limiting distances are greater than the typical scale height of AGB stars of 100~pc.}
one can estimate that the  'high' tier represents about 1\% of all AGB stars, while the  'extreme' tier is 20 times less abundant.

\begin{figure}
\centering

\includegraphics[width=0.85\hsize]{TD_V_Per.ps}

\includegraphics[width=0.85\hsize]{TD_V_Lum.ps}

\includegraphics[width=0.85\hsize]{TD_Mdot_Lum.ps}

\includegraphics[width=0.85\hsize]{TD_Mdot_V.ps}

\caption[]{ 
  From top to bottom: $V_{\rm exp}$ plotted against pulsation period and luminosity,
        \mdot\ (the geometric mean from columns~11--13 in Table~\ref{Tab-SED}) plotted against luminosity and $V_{\rm exp}$. 
      Plotted are the  C stars in the LMC (red-filled squares) and in the SMC (pink-filled triangles),
  Galactic C stars from \citet{SO2001}, \citet{RO14}, and \citet{Danilovich15} 
  (open black squares), and the highly red Galactic C stars (black-filled squares).
  Objects with very large error bars in the expansion velocity ($>$2.4~\ks\ in the SMC, $>$3.0~\ks\ in the LMC) are not plotted.
  The blue symbols represent models selected from \cite{Bladh19} (see text) and indicate the SMC (blue-filled triangles), LMC (blue-filled squares), 
  and MW (open blue squares). The SMC and MW sources are offset by $-0.01$ and $+0.01$~dex in $\log L$ for clarity.
} 

\label{Fig-Comp} 
\end{figure}

\begin{figure}
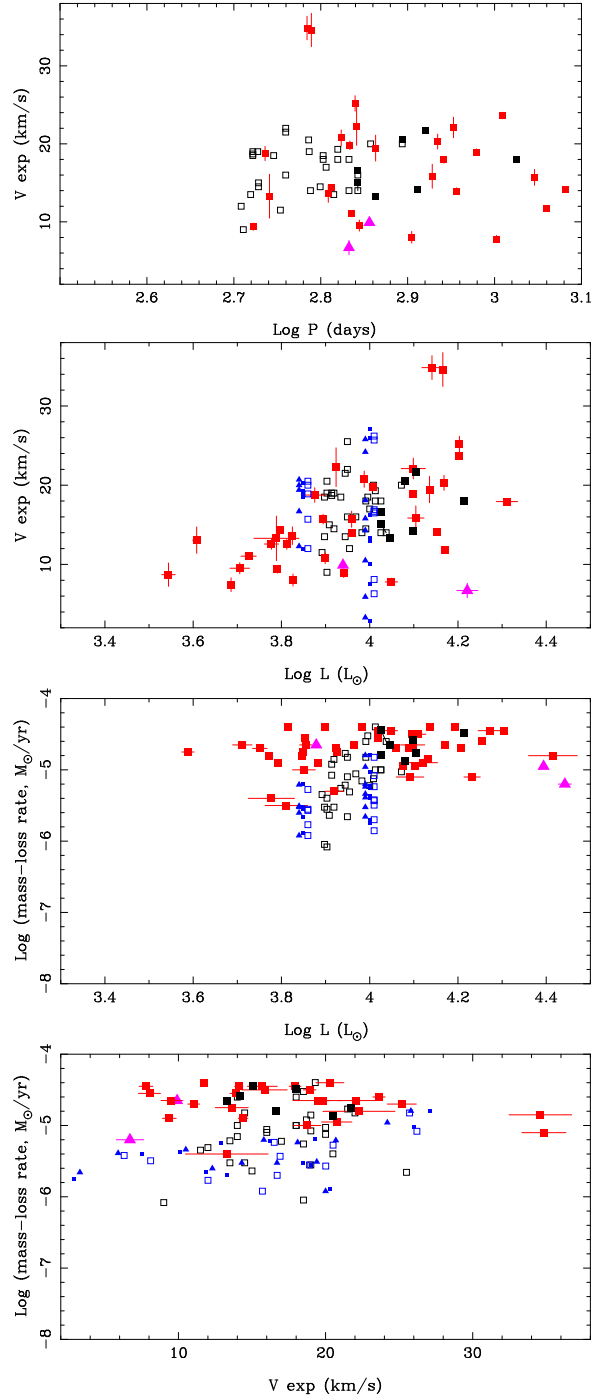

\centering

\includegraphics[width=0.85\hsize]{TD_V_Per_500.ps}

\includegraphics[width=0.85\hsize]{TD_V_Lum_500.ps}

\includegraphics[width=0.85\hsize]{TD_Mdot_Lum_500.ps}

\includegraphics[width=0.85\hsize]{TD_Mdot_V_500.ps}

\caption[]{ 
  As Figure~\ref{Fig-Comp}, but with the pulsation period limited to $>~500$~days. 
} 

\label{Fig-CompP} 
\end{figure}

\section{Summary and conclusions}

Our observations present the first detection of CO in SMC AGB stars (in two out of three targets) and extend the detections in LMC AGB carbon stars from 4 to 37.
We reported the first detection of $^{13}$CO for a LMC carbon star, IRAS 05568.

Radiation-driven wind theory predicts $V_{\rm exp} \sim$GTD$^{-0.5}$ (e.g. \citealt{HTT94}).
The median GTD  is derived to be 1500 for the three SMC stars, 400 for the 42 LMC stars, and 190 for the seven Galactic stars. 
Between SMC and LMC, one would therefore expect a difference in expansion velocities by a factor of 1.9.
The two stars detected in the SMC have among the lowest expansion velocities -- 6.7~\ks\ and 9.9~\ks -- while the
median among the LMC detections is 15.7~\ks. This is indeed close to a factor of two, but only based on a very small sample.
The difference in the GTD ratio between LMC and the MW implies a  difference in expansion velocity by a factor of about 1.4, which
is not supported  by the present data.
Interestingly, a few LMC objects have expansion velocities above 30~\ks\ (31-35~\ks).
No counterparts exist in the Galactic samples studied here, although such carbon stars do exist. For example, 13 sources among the 331 C stars
detected in CO in \cite{Groenewegen2002b} have expansion velocities ranging from 31 to 43~\ks.

Extending the sample further to better probe the range of intermediate luminosities at slightly lower MLRs is a possibility.
This ALMA programme has demonstrated that CO line intensities can be predicted sufficiently accurately to ensure a high detection rate.
Such a programme would require an order of magnitude more observing time (approximately 100 hours) for a dozen objects.
One approach we want to pursue in the future is to obtain observations at other transitions of selected stars and model the line shapes
and intensities with a RT code. This would constrain the CO emission region (and the role of photodissociation) and lead to a more accurate
estimate of the gas MLR.

\section*{Data availability}

The complete set of figures shown in Appendix~\ref{App-Mom0}, \ref{App-LP}, \ref{App-Periods}, and \ref{App-SED}
is available at \protect\url{https://doi.org/10.5281/zenodo.21700114}.

\begin{acknowledgements} 
This paper makes use of the following ALMA data: ADS/JAO.ALMA\#2023.1.00681.S.
ALMA is a partnership of ESO
(representing its member states), NSF (USA) and NINS (Japan), together
with NRC (Canada) and NSC and ASIAA (Taiwan) and KASI (Republic of Korea), 
in cooperation with the Republic of Chile. 
The Joint ALMA Observatory is operated by ESO, AUI/NRAO and NAOJ.
MG acknowledges assistance from Allegro, the ALMA Regional Centre node in the Netherlands, and in particular Megan Lewis.
AN acknowledges support from the Narodowe Centrum Nauki (NCN), Poland, through the SONATA BIS grant UMO-2020/38/E/ST9/00077.
R.S.’s contribution to the research described here was carried out at the Jet Propulsion Laboratory, California Institute of Technology,
under a contract with NASA (80NM0018D0004), and funded in part by NASA via ADAP awards.
L.~Decin acknowledge support from the FWO grant G0B3823N, the  FWO grant G099720N, the KU Leuven C1 excellence grant MAESTRO C16/17/007,
and the KU Leuven Methusalem SOUL grant METH/24/012.
KJ acknowleges the support of Swedish National Space Agency.

\end{acknowledgements}

\bibliographystyle{aa.bst}
\bibliography{references.bib}

\clearpage
\onecolumn
\begin{appendix}
\nolinenumbers

\section{Image cut-outs}
\label{App-Mom0}

 \begin{figure*}[h]
  \centering

\begin{minipage}{0.49\textwidth}
\resizebox{\hsize}{!}{\includegraphics[angle=0]{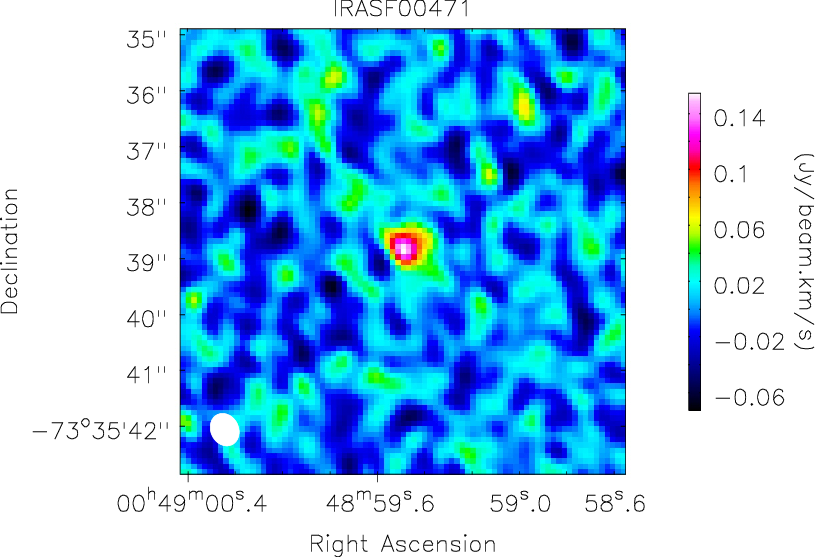}} 
\end{minipage}
\begin{minipage}{0.49\textwidth}
\resizebox{\hsize}{!}{\includegraphics[angle=0]{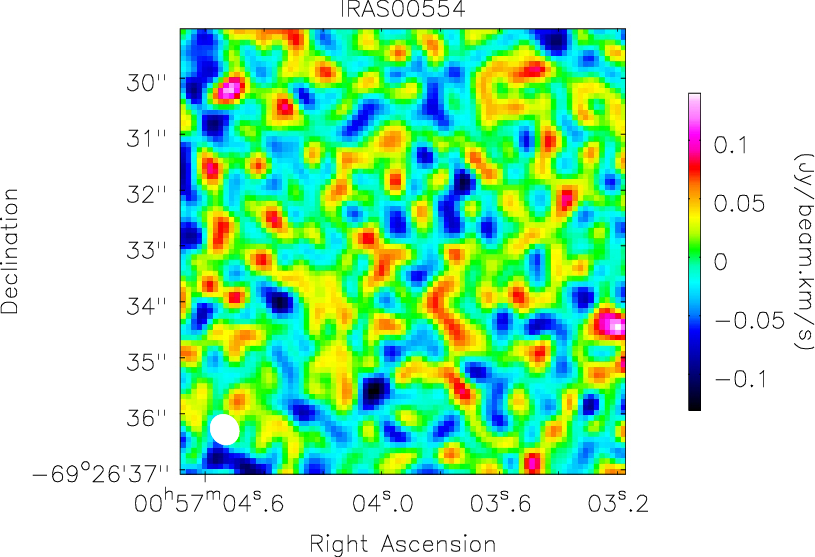}} 
\end{minipage}
 
\begin{minipage}{0.49\textwidth}
\resizebox{\hsize}{!}{\includegraphics[angle=0]{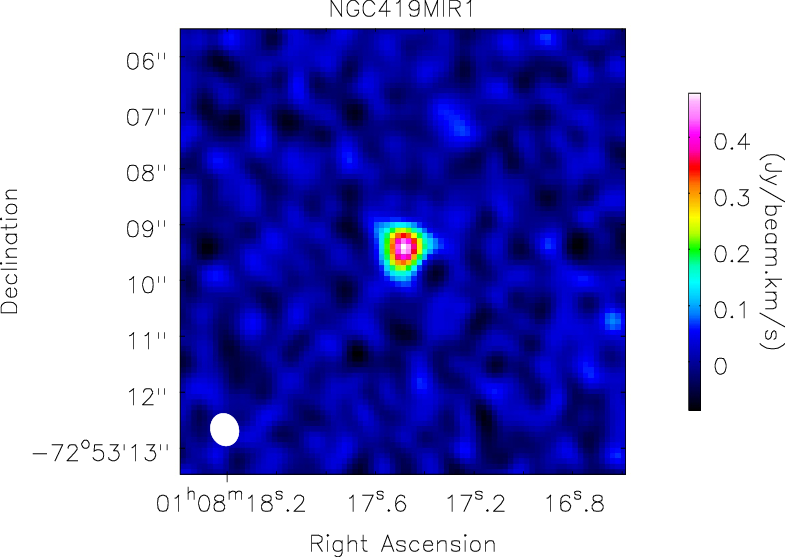}} 
\end{minipage}
\begin{minipage}{0.49\textwidth}
\resizebox{\hsize}{!}{\includegraphics[angle=0]{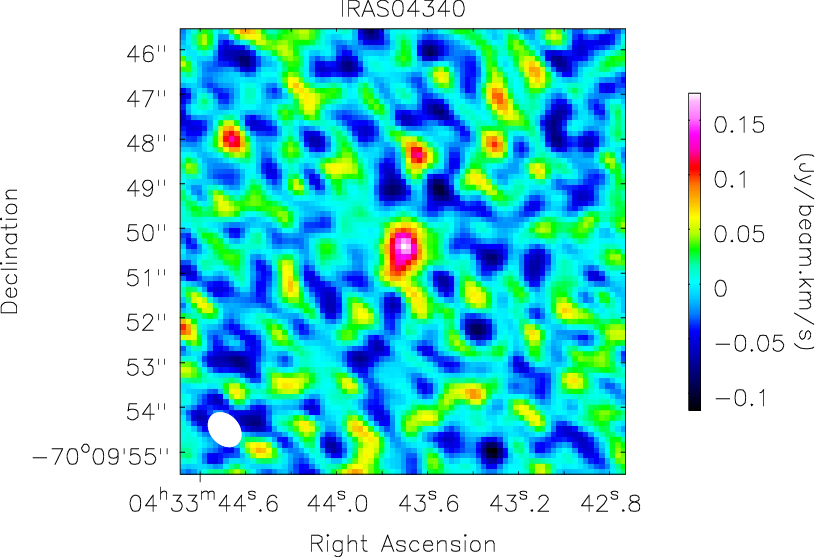}} 
\end{minipage}
 
\begin{minipage}{0.49\textwidth}
\resizebox{\hsize}{!}{\includegraphics[angle=0]{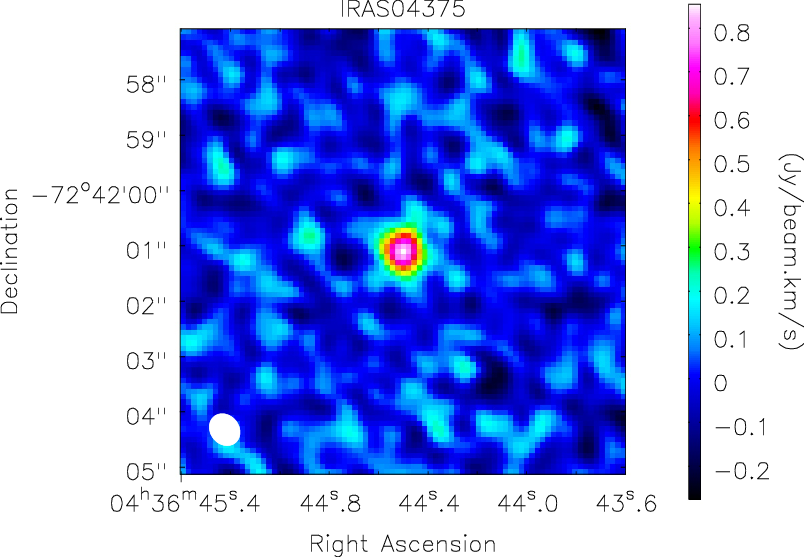}} 
\end{minipage}
\begin{minipage}{0.49\textwidth}
\resizebox{\hsize}{!}{\includegraphics[angle=0]{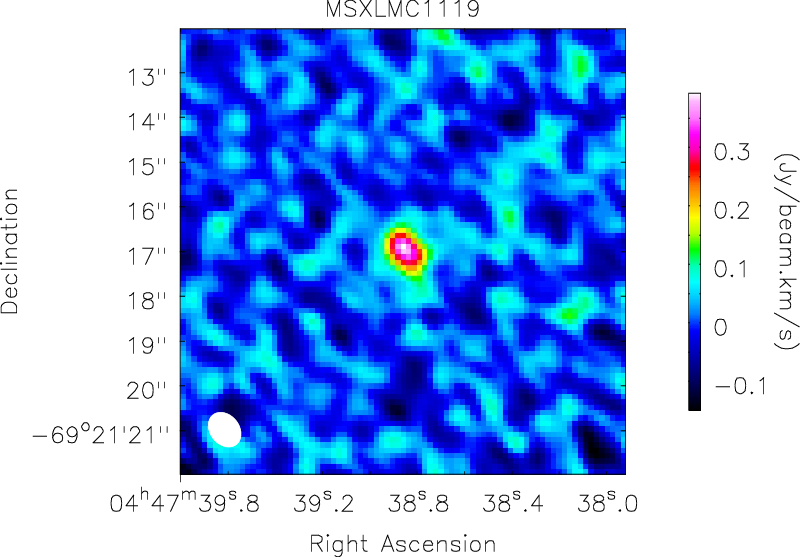}} 
\end{minipage}

\caption{\label{App-Fig-CO} Example of 10\arcsec$\times$10\arcsec\ zero-moment maps of the $^{12}$CO emission centred on the source.
  The white ellipse in the lower left corners represents the beam. The complete set of figures is available via Zenodo.}
\end{figure*}
 
\clearpage

\section{CO line profiles}
\label{App-LP}

\begin{figure*}[h]
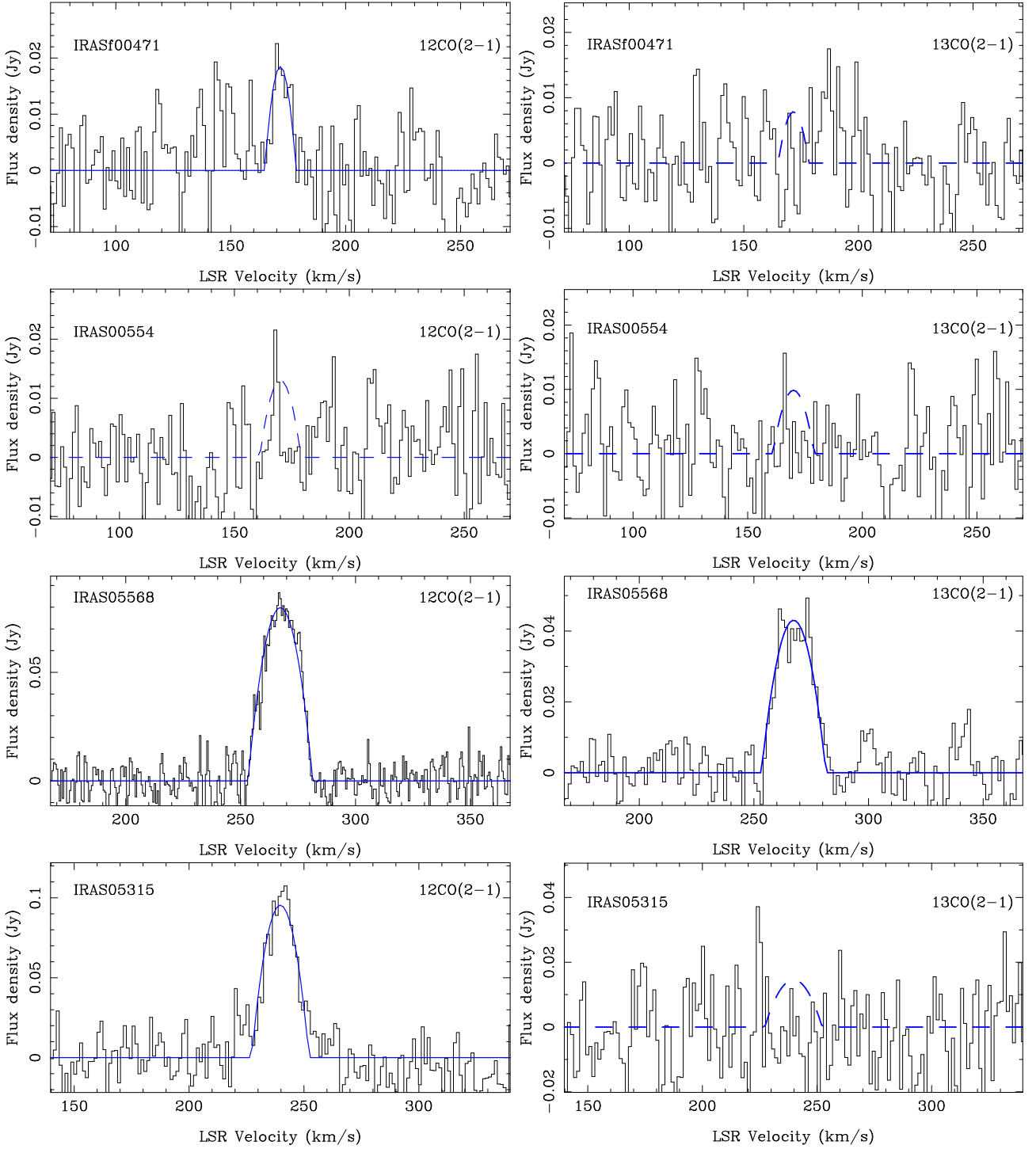

  \centering
  
\begin{minipage}{0.46\textwidth}
\resizebox{\hsize}{!}{\includegraphics[angle=0]{profilefit_IRASf00471.ps}} 
\end{minipage}
\begin{minipage}{0.46\textwidth}
\resizebox{\hsize}{!}{\includegraphics[angle=0]{profilefit13CO_IRASf00471.ps}} 
\end{minipage}
 
\begin{minipage}{0.46\textwidth}
\resizebox{\hsize}{!}{\includegraphics[angle=0]{profilefit_IRAS00554.ps}} 
\end{minipage}
\begin{minipage}{0.46\textwidth}
\resizebox{\hsize}{!}{\includegraphics[angle=0]{profilefit13CO_IRAS00554.ps}} 
\end{minipage}

\begin{minipage}{0.46\textwidth}
\resizebox{\hsize}{!}{\includegraphics[angle=0]{profilefit_IRAS05568.ps}} 
\end{minipage}
\begin{minipage}{0.46\textwidth}
\resizebox{\hsize}{!}{\includegraphics[angle=0]{profilefit13CO_IRAS05568.ps}} 
\end{minipage}
 
\begin{minipage}{0.46\textwidth}
\resizebox{\hsize}{!}{\includegraphics[angle=0]{profilefit_IRAS05315.ps}} 
\end{minipage}
\begin{minipage}{0.46\textwidth}
\resizebox{\hsize}{!}{\includegraphics[angle=0]{profilefit13CO_IRAS05315.ps}} 
\end{minipage}

\caption{\label{App-Fig-LP}Examples of $^{12}$CO(2--1) and $^{13}$CO(2--1) line profiles.
The best ﬁtting models are shown as blue full lines, and the 3$\sigma$ upper limits in the case of non-detections (see text) are shown as dashed lines.
The complete set of figures is available via Zenodo
}
\end{figure*}

\begin{multicols}{2}
\noindent
  Figure~\ref{App-Fig-LP} shows the $^{12}$CO(2--1) and $^{13}$CO(2--1) line profiles.
The blue lines refer to the best fitting models.
In case of a $^{13}$CO non-detection a model representing the 3$\sigma$ upper limit on the integrated intensity is shown, plotted at the systemic
velocity of the $^{12}$CO emission and for a fixed expansion velocity.
In case of a  $^{12}$CO non-detection a model representing the 3$\sigma$ upper limit on the integrated intensity is shown
for a random systemic velocity and a fixed expansion velocity of 15~\ks\ (9~\ks\ for the SMC source).
The complete set of figures is available via Zenodo.
\end{multicols}

\clearpage

\section{Period analysis}
\label{App-Periods}

\begin{figure*}
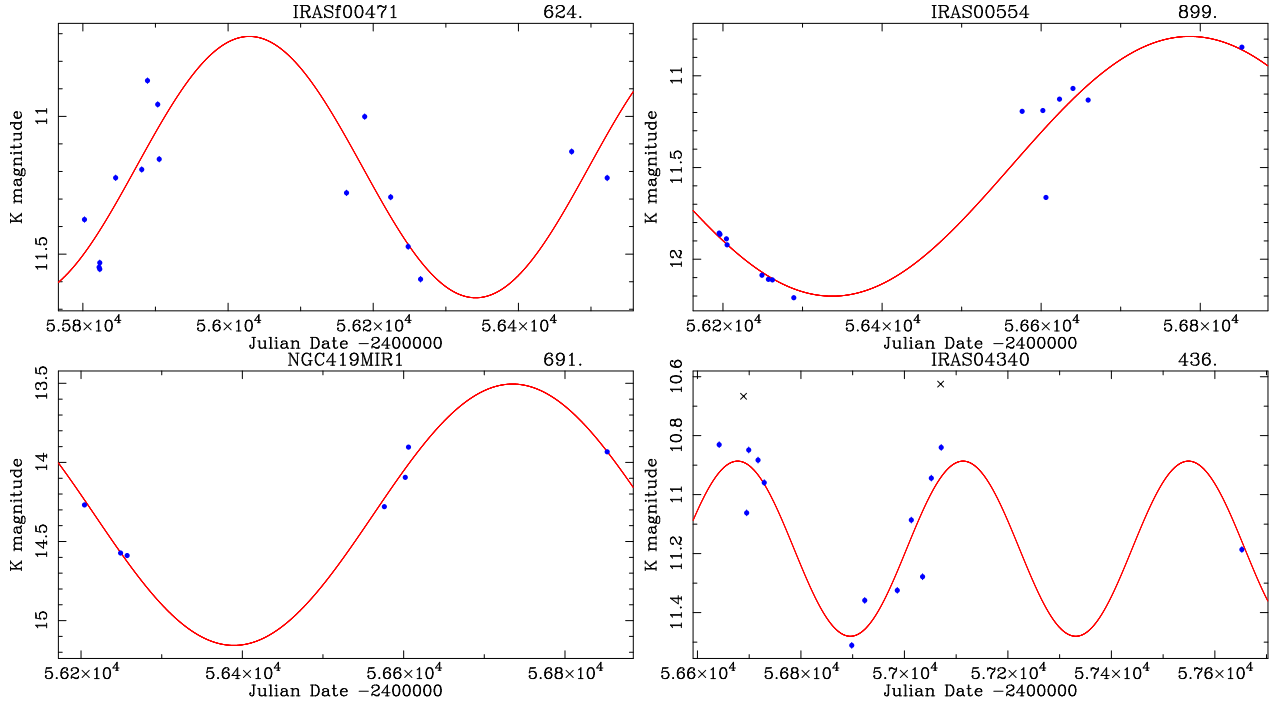

   \centering
   
\begin{minipage}{0.45\textwidth}
\resizebox{\hsize}{!}{\includegraphics[angle=0]{12.247879_-73.594121.ps}} 
\end{minipage}
\begin{minipage}{0.45\textwidth}
\resizebox{\hsize}{!}{\includegraphics[angle=0]{14.266367_-73.587387.ps}} 
\end{minipage}

\begin{minipage}{0.45\textwidth}
\resizebox{\hsize}{!}{\includegraphics[angle=0]{17.072829_-72.885948.ps}} 
\end{minipage}
\begin{minipage}{0.45\textwidth}
\resizebox{\hsize}{!}{\includegraphics[angle=0]{68.432358_-70.164048.ps}} 
\end{minipage}

\caption{\label{AppFig-VMC} Examples of the LC and the model fit (in red) in the $K$-band.
  A cross indicates a data point is excluded from the fit.
  The source name and period are listed at the top of the plot.
  The complete set of figures is available via Zenodo.}
 \end{figure*}

\begin{figure*}
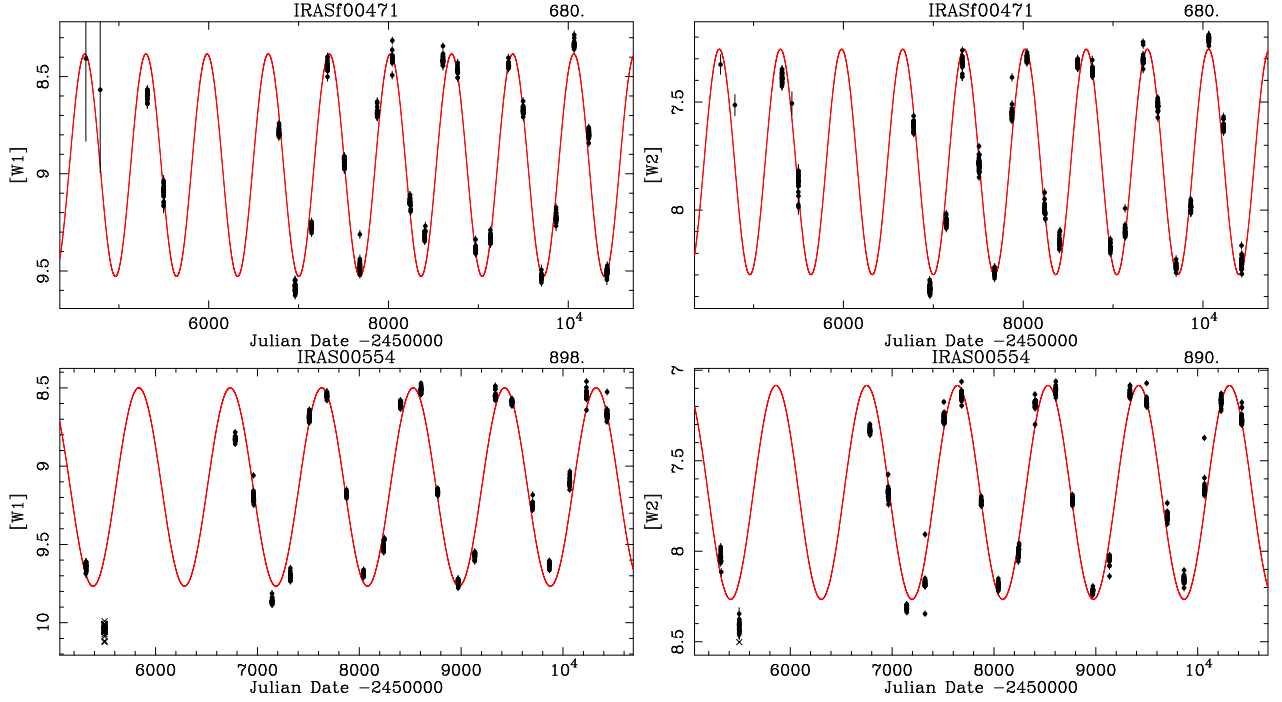


   \centering

   \begin{minipage}{0.45\textwidth}
\resizebox{\hsize}{!}{\includegraphics[angle=-0]{12.247879_-73.594120_W1.ps}} 
\end{minipage}
\begin{minipage}{0.45\textwidth}
\resizebox{\hsize}{!}{\includegraphics[angle=-0]{12.247879_-73.594120_W2.ps}} 
\end{minipage}
 
\begin{minipage}{0.45\textwidth}
\resizebox{\hsize}{!}{\includegraphics[angle=-0]{14.266367_-73.587387_W1.ps}} 
\end{minipage}
\begin{minipage}{0.45\textwidth}
\resizebox{\hsize}{!}{\includegraphics[angle=-0]{14.266367_-73.587387_W2.ps}} 
\end{minipage}

\caption{\label{AppFig-WISE} Examples of the LC and the model fit (in red) in the W1-band (left panel), and W2-band (right panel)
  for the same stars as in Fig.~\ref{AppFig-VMC}.
  The complete set of figures is available via Zenodo.}
 \end{figure*}

\begin{multicols}{2}
\noindent
  Tables~\ref{Tab-VMC-Periods} and \ref{Tab-WISE-Periods} list the derived periods based on VMC and WISE data, respectively.
Figures~\ref{AppFig-VMC}     and \ref{AppFig-WISE} show examples of the LCs and the fitted models.
\end{multicols}

\begin{table*}[h]

  \fontsize{8.5}{10.2}\selectfont
  
  \setlength{\tabcolsep}{1.1mm}
\caption{\label{Tab-VMC-Periods} Period analysis in the $K$-band.}
\begin{tabular}{lcccccrcccrcrrclccrlllllllllllll}
\hline
\hline
\centering

Name & Ra$_i$     & Dec$_i$  & Ra$_f$ & Dec$_f$ & Dist &  N & $P_{\rm K}$ & $\sigma_{\rm P_K}$ &  AmpK & $\sigma_{\rm AmpK}$ &  $K$ & $\sigma_{\rm K}$ & $\chi_{\rm K}^2$    \\
     &  (deg) & (deg) & (deg) & (deg) &   (\arcsec) & & (d)       &  (d)              &  (mag) & (mag)            & (mag) & (mag)          &              \\
\hline

IRASf00471 &  12.247879  & -73.594121  &   12.247876  & -73.594104  & 0.06  & 16 & 623.6 &  27.1 & 0.47 & 0.10 & 11.184 &  0.052 &  244.2  \\
IRAS00554 &  14.266367  & -73.587387  &   14.266429  & -73.587419  & 0.13  & 15 & 898.8 &  84.8 & 0.71 & 0.15 & 11.493 &  0.065 &  207.8  \\
NGC419MIR1 &  17.072829  & -72.885948  &   17.072880  & -72.885959  & 0.06  &  7 & 691.2 &  22.7 & 0.83 & 0.16 & 14.331 &  0.056 &   53.9  \\
IRAS04340 &  68.432358  & -70.164048  &   68.431999  & -70.164004  & 0.47  & 15 & 435.7 &  16.0 & 0.30 & 0.07 & 11.183 &  0.036 &  109.9  \\
IRAS04375 &  69.185429  & -72.700292  &   69.185315  & -72.700325  & 0.17  & 15 & 746.5 &  78.8 & 0.66 & 0.10 & 12.192 &  0.022 &   48.9  \\
MSXLMC1119 &  71.911867  & -69.354717  &   71.911847  & -69.354718  & 0.03  & 16 & 488.5 &   5.3 & 0.49 & 0.03 & 11.571 &  0.016 &   32.9  \\
IRAS04496 &  72.326917  & -69.887359  &   72.327053  & -69.887356  & 0.17  & 12 &      &      &      &     & 10.983 &  0.022 &   43.9  \\
IRAS04523 &  72.962163  & -70.642241  &   72.962073  & -70.642208  & 0.16  & 18 & 857.4 &  42.9 & 0.45 & 0.07 & 11.365 &  0.083 &  204.7  \\
SAGEMCJ045344 &  73.434475  & -66.196132  &   73.434395  & -66.196116  & 0.13  & 15 & 762.9 &  47.9 & 0.89 & 0.16 & 15.520 &  0.030 &    9.4  \\
IRAS04557 &  73.912421  & -67.819591  &   73.912385  & -67.819623  & 0.12  & 15 & 651.3 &   2.9 & 0.69 & 0.01 & 12.839 &  0.004 &    1.1  \\
MSXLMC1220 &  73.924079  & -68.956284  &   73.924134  & -68.956312  & 0.12  & 14 & 433.9 &  31.5 & 0.42 & 0.12 & 11.337 &  0.105 &  456.7  \\
MSXLMC1303 &  74.675175  & -68.120835  &   74.675133  & -68.120909  & 0.27  & 16 & 1184.2 & 852.1 & 0.28 & 0.21 & 15.765 &  0.194 &  375.3  \\
MSXLMC1282 &  75.253751  & -67.589911  &   75.253617  & -67.589880  & 0.21  & 16 & 498.4 &  27.5 & 0.43 & 0.09 & 11.379 &  0.063 &  461.2  \\
ERO0502315 &  75.631267  & -68.093301  &   75.631594  & -68.093646  & 1.32  & 1 &      &      &      &      &        &        &         \\
MSXLMC91 &  75.910721  & -68.886864  &   75.910566  & -68.886888  & 0.22  & 14 & 889.4 &  13.9 & 0.80 & 0.03 & 13.002 &  0.014 &   17.7  \\
TRM74 &  75.929217  & -66.749036  &   75.929193  & -66.749080  & 0.17  & 4 &      &      &      &      &        &        &         \\
ERO0504056 &  76.023421  & -68.394531  &   76.024634  & -68.394107  & 2.22  & 10 &      &      &      &     & 19.168 &  0.061 &    0.5  \\
MSXLMC87 &  77.581901  & -69.830974  &   77.581756  & -69.830970  & 0.18  & 14 & 613.8 & 789.8 & 0.23 & 0.52 & 11.078 &  0.440 &  119.9  \\
IRAS05113 &  77.736892  & -69.591809  &   77.736634  & -69.591819  & 0.33  & 14 & 756.7 & 1029.5 & 0.36 & 0.35 & 11.173 &  0.296 &  845.7  \\
IRAS05132 &  78.212758  & -69.630653  &   78.212652  & -69.630661  & 0.14  & 14 & 697.6 &  11.3 & 0.65 & 0.01 & 12.433 &  0.006 &    1.1  \\
IRAS05133 &  78.257442  & -69.564153  &   78.257558  & -69.564206  & 0.24  & 14 &      &      &      &     & 16.239 &  0.006 &    0.2  \\
ERO0518484 &  79.701683  & -69.559625  &   79.701263  & -69.559886  & 1.08  & 30 &      &      &      &     & 16.648 &  0.010 &    3.2  \\
IRAS05190 &  79.734463  & -67.751243  &   79.734423  & -67.751236  & 0.06  & 14 & 894.1 &  23.7 & 0.98 & 0.02 & 13.848 &  0.016 &   14.2  \\
TRM88 &  80.080875  & -66.596626  &   80.080755  & -66.596613  & 0.18  & 16 &      &      &      &     & 10.704 &  0.034 &  169.4  \\
MSXLMC527 &  80.581604  & -65.722303  &   80.582272  & -65.721907  & 1.74  & 16 & 1068.9 & 221.1 & 0.37 & 0.09 & 13.600 &  0.076 &  205.9  \\
ERO0525406 &  81.419396  & -70.140901  &   81.419323  & -70.141075  & 0.63  & 15 &      &      &      &     & 17.126 &  0.010 &    0.6  \\
MSXLMC474 &  81.466101  & -68.776167  &   81.466071  & -68.776204  & 0.14  & 15 & 640.7 &  62.4 & 0.72 & 0.09 & 11.635 &  0.034 &   60.3  \\
IRAS05315 &  82.683892  & -71.716808  &   82.683878  & -71.716831  & 0.08  & 29 &      &      &      &     & 17.029 &  0.014 &    4.7  \\
MSXLMC1780 &  83.257483  & -68.399565  &   83.257376  & -68.399582  & 0.15  & 17 & 827.2 &   6.2 & 1.00 & 0.04 & 13.452 &  0.022 &   20.2  \\
IRAS05373 &  84.100671  & -72.692391  &   84.100610  & -72.692374  & 0.08  & 15 & 812.9 &  37.8 & 0.63 & 0.11 & 12.659 &  0.080 &  178.2  \\
MSXLMC971 &  84.966001  & -70.021426  &   84.965917  & -70.021386  & 0.18  & 14 & 619.9 &  24.8 & 0.96 & 0.02 & 13.503 &  0.018 &    4.3  \\
MSXLMC937 &  85.150438  & -69.880528  &   85.150268  & -69.880480  & 0.27  & 14 & 642.5 &  93.2 & 0.70 & 0.04 & 11.358 &  0.054 &   26.0  \\
IRAS05416 &  85.336488  & -69.078812  &   85.336413  & -69.078815  & 0.11  & 15 & 892.8 &  20.8 & 0.92 & 0.28 & 17.726 &  0.055 &    0.3  \\
LMCLPV76711 &  85.571808  & -70.538971  &   85.571700  & -70.539003  & 0.18  & 17 & 515.6 &  23.7 & 0.36 & 0.11 & 11.918 &  0.039 &  160.2  \\
IRAS05495 &  87.250017  & -70.556247  &   87.250043  & -70.556172  & 0.27  & 12 &      &      &      &     & 19.167 &  0.059 &    0.2  \\
ERO0550261 &  87.608879  & -69.934185  &   87.609920  & -69.934187  & 1.29  &  8 &      &      &      &     & 19.254 &  0.074 &    0.5  \\
IRAS05515 &  87.708112  & -71.393213  &   87.708039  & -71.393240  & 0.13  & 15 & 643.9 &  10.5 & 0.59 & 0.04 & 13.212 &  0.016 &   23.9  \\
MSXLMC1797 &  87.889933  & -71.326194  &   87.887844  & -71.325661  & 3.08  & 1 &      &      &      &      &        &        &         \\
IRAS05568 &  89.161538  & -67.892749  &   89.161484  & -67.892776  & 0.13  & 11 & 1118.7 &   4.7 & 1.48 & 0.04 & 17.458 &  0.025 &    0.4  \\
IRAS06028 &  90.688067  & -67.378653  &   90.688066  & -67.378676  & 0.08  & 14 & 871.2 &   9.5 & 1.00 & 0.05 & 13.454 &  0.016 &   25.3  \\
IRAS06108 &  92.544704  & -70.767581  &   92.544461  & -70.767573  & 0.29  & 26 & 530.8 &   2.1 & 0.70 & 0.03 & 12.954 &  0.009 &   13.5  \\
IRAS05506 &  87.485502  & -70.886597  &   87.485612  & -70.886609  & 0.14  & 15 & 1310.5 &  64.4 & 0.93 & 0.04 & 12.429 &  0.046 &    5.5  \\
IRAS05125 &  78.003364  & -70.540098  &   78.003223  & -70.540034  & 0.29  & 16 & 1347.9 &  44.1 & 0.46 & 0.02 & 15.289 &  0.013 &   11.8  \\
ERO0529379 &  82.407854  & -72.831365  &   82.406845  & -72.831189  & 2.10  & 30 &      &      &      &     & 16.806 &  0.056 &  102.2  \\
ERO0518117 &  79.548769  & -70.507508  &   79.549041  & -70.507507  & 0.33  &  5 &      &      &      &     & 18.975 &  0.079 &    0.8  \\

\hline
\end{tabular}
\tablefoot{
  Listed are the source name, the input coordinates used to search the VMC archive, the "final" coordinates being the average of the coordinates
  of all entries contributing to the LC, the distance between input and final coordinates, the number of data points,
  period, amplitude, and mean magnitude (with errors), and the reduced $\chi^2$. All error bars have been scaled to give a reduced $\chi^2$ of unity.
  \\
}\\

\vfill
\end{table*}

\begin{sidewaystable*}

  \fontsize{7.0}{8.5}\selectfont
  
  \setlength{\tabcolsep}{1.7mm}
\caption{\label{Tab-WISE-Periods} Period analysis in the WISE bands.}
\begin{tabular}{lcccccrrrrccccrccccrllll}
\hline
\hline
\centering

Name & Ra$_i$ & Dec$_i$  & Ra$_f$ & Dec$_f$ & Dist &  $P1$ & $\sigma_{\rm P1}$ & $P2$ & $\sigma_{\rm P2}$ &  Amp1 & $\sigma_{\rm A1}$ &  $W1$ & $\sigma_{\rm W1}$ & $\chi_{\rm r,W1}^2$ & Amp2 & $\sigma_{\rm A2}$ &  $W2$ & $\sigma_{\rm W2}$ & $\chi_{\rm r,W2}^2$  \\
     & (deg) & (deg)     & (deg) & (deg)   & (\arcsec) & (d) &  (d)          & (d) & (d)               & (mag) & (mag) & (mag) & (mag) &   & (mag) & (mag) &  (mag) & (mag) &  \\
\hline

IRASf00471 &  12.247879  & -73.594120  &  12.247892  & -73.594114  & 0.07  &  680 &  1.0 &  680 &  0.8 & 0.57 & 0.01 &  8.956 &  0.004 &   26.4  & 0.52 & 0.01 &  7.777 &  0.003 &   22.4 \\
IRAS00554 &  14.266367  & -73.587387  &  14.266423  & -73.587432  & 0.18  &  898 &  1.1 &  890 &  0.9 & 0.63 & 0.02 &  9.133 &  0.004 &   33.9  & 0.59 & 0.01 &  7.675 &  0.004 &   32.1 \\
NGC419MIR1 &  17.072829  & -72.885948  &  17.073007  & -72.885810  & 0.53  &  717 &  1.3 &  718 &  0.6 & 0.40 & 0.02 &  9.881 &  0.012 &   32.7  & 0.53 & 0.01 &  8.723 &  0.007 &   13.6 \\
IRAS04340 &  68.432358  & -70.164048  &  68.432030  & -70.164015  & 0.42  &  439 &  0.5 &  440 &  0.4 & 0.51 & 0.02 &  9.740 &  0.005 &   62.0  & 0.46 & 0.01 &  8.701 &  0.004 &   42.2 \\
IRAS04375 &  69.185429  & -72.700292  &  69.185376  & -72.700302  & 0.11  &  692 &  0.6 &  695 &  0.5 & 0.60 & 0.01 &  9.719 &  0.004 &   27.2  & 0.52 & 0.01 &  8.091 &  0.003 &   22.0 \\
MSXLMC1119 &  71.911867  & -69.354717  &  71.911858  & -69.354727  & 0.09  &  545 &  0.3 &  544 &  0.3 & 0.47 & 0.01 &  9.468 &  0.002 &   11.2  & 0.44 & 0.01 &  8.227 &  0.002 &   12.6 \\
IRAS04496 &  72.326917  & -69.887359  &  72.327036  & -69.887369  & 0.16  &  824 &  3.7 &  816 &  4.6 & 0.12 & 0.01 &  8.029 &  0.004 &   23.7  & 0.10 & 0.01 &  7.027 &  0.004 &   35.0 \\
IRAS04523 &  72.962163  & -70.642240  &  72.962078  & -70.642242  & 0.13  &  875 &  0.8 &  874 &  0.6 & 0.70 & 0.01 &  8.940 &  0.003 &   28.8  & 0.64 & 0.01 &  7.428 &  0.003 &   21.4 \\
SAGEMCJ045344 &  73.434475  & -66.196132  &  73.434463  & -66.196115  & 0.13  &  761 &  0.5 &  761 &  0.4 & 0.78 & 0.01 & 11.168 &  0.003 &   20.7  & 0.70 & 0.01 &  8.844 &  0.002 &   18.4 \\
IRAS04557 &  73.912421  & -67.819591  &  73.912433  & -67.819599  & 0.11  &  728 &  0.6 &  732 &  0.4 & 0.70 & 0.01 & 10.078 &  0.003 &   32.9  & 0.59 & 0.01 &  8.266 &  0.002 &   19.7 \\
MSXLMC1220 &  73.924079  & -68.956284  &  73.924132  & -68.956273  & 0.11  &  513 &  0.5 &  514 &  0.3 & 0.40 & 0.01 &  9.264 &  0.004 &   45.0  & 0.33 & 0.01 &  8.062 &  0.002 &   20.7 \\
MSXLMC1303 &  74.675175  & -68.120835  &  74.675107  & -68.120882  & 0.23  & 1005 &  4.5 & 1005 &  4.4 & 0.49 & 0.03 & 11.086 &  0.011 &  329.9  & 0.42 & 0.03 &  8.575 &  0.009 &  359.5 \\
MSXLMC1282 &  75.253750  & -67.589910  &  75.253733  & -67.589893  & 0.10  &  608 &  0.7 &  609 &  0.7 & 0.53 & 0.01 &  8.930 &  0.004 &   62.7  & 0.42 & 0.02 &  7.629 &  0.004 &   77.0 \\
ERO0502315 &  75.631267  & -68.093300  &  75.631233  & -68.093309  & 0.37  &      &      & 15938 & 449.4 &      &     & 15.753 &  0.025 &    1.9  & 0.80 & 0.03 & 12.547 &  0.032 &    3.7 \\
MSXLMC91 &  75.910721  & -68.886864  &  75.910640  & -68.886878  & 0.15  &  897 &  0.8 &  897 &  0.7 & 0.77 & 0.02 &  9.624 &  0.003 &   30.5  & 0.66 & 0.01 &  7.740 &  0.002 &   26.7 \\
TRM74 &  75.929217  & -66.749036  &  75.929257  & -66.749079  & 0.32  & 8092 & 3137 & 10256 & 1631 & 0.04 & 0.02 & 15.031 &  0.018 &    2.2  & 0.05 & 0.01 & 12.513 &  0.010 &    2.3 \\
ERO0504056 &  76.023421  & -68.394531  &  76.023374  & -68.394518  & 0.44  &      &      & 8920 & 277.1 &      &      &        &        &         & 0.29 & 0.01 & 13.329 &  0.006 &    2.0 \\
MSXLMC87 &  77.581900  & -69.830974  &  77.581757  & -69.830969  & 0.18  &  617 &  1.7 &  614 &  1.6 & 0.37 & 0.03 &  8.870 &  0.008 &  251.9  & 0.34 & 0.01 &  7.662 &  0.007 &  255.9 \\
IRAS05113 &  77.736892  & -69.591809  &  77.736743  & -69.591808  & 0.21  &  633 &  1.1 &  629 &  0.9 & 0.65 & 0.03 &  9.156 &  0.006 &  123.3  & 0.52 & 0.02 &  7.735 &  0.004 &   73.9 \\
IRAS05132 &  78.212758  & -69.630653  &  78.212667  & -69.630640  & 0.15  &  683 &  0.7 &  681 &  0.5 & 0.63 & 0.01 &  9.452 &  0.003 &   38.8  & 0.58 & 0.01 &  7.928 &  0.002 &   22.9 \\
IRAS05133 &  78.257442  & -69.564153  &  78.257370  & -69.564149  & 0.51  &      &      &  702 &  6.8 &      &     & 16.282 &  0.209 &    0.0  & 0.08 & 0.02 & 13.869 &  0.007 &    1.7 \\
ERO0518484 &  79.701683  & -69.559625  &  79.701599  & -69.559559  & 0.26  &      &      &  803 & 22.3 &      &     & 15.307 &  0.070 &    0.3  & 0.21 & 0.03 & 12.719 &  0.031 &    5.0 \\
IRAS05190 &  79.734463  & -67.751243  &  79.734384  & -67.751229  & 0.15  &  960 &  1.6 &  950 &  1.1 & 0.96 & 0.02 & 10.293 &  0.006 &  154.4  & 0.77 & 0.01 &  8.081 &  0.004 &  106.9 \\
TRM88 &  80.080875  & -66.596626  &  80.080814  & -66.596624  & 0.12  &  538 &  0.7 &  536 &  0.5 & 0.41 & 0.01 &  8.985 &  0.005 &  118.3  & 0.38 & 0.01 &  7.950 &  0.003 &   76.7 \\
MSXLMC527 &  80.581604  & -65.722303  &  80.582113  & -65.722210  & 0.77  &      &      &      &      &      &     & 11.300 &  0.212 &  591.2  &      &     &  9.294 &  0.192 &  813.8 \\
ERO0525406 &  81.419396  & -70.140901  &  81.419409  & -70.140876  & 0.09  &      &      &  828 & 27.8 &      &     & 15.703 &  0.351 &    4.4  & 0.26 & 0.13 & 13.297 &  0.087 &    3.1 \\
MSXLMC474 &  81.466100  & -68.776167  &  81.466072  & -68.776173  & 0.10  &  617 &  0.8 &  625 &  0.5 & 0.55 & 0.02 &  9.098 &  0.004 &   83.5  & 0.51 & 0.01 &  7.653 &  0.002 &   41.0 \\
IRAS05315 &  82.683892  & -71.716808  &  82.684316  & -71.716749  & 0.54  & 3111 & 199.8 & 2985 & 54.4 & 0.02 & 0.01 & 13.831 &  0.003 &    2.5  & 0.04 & 0.01 & 11.980 &  0.002 &    2.3 \\
MSXLMC1780 &  83.257483  & -68.399565  &  83.257332  & -68.399570  & 0.22  &  848 &  0.8 &  848 &  0.7 & 0.76 & 0.01 & 10.304 &  0.003 &   67.9  & 0.69 & 0.01 &  8.240 &  0.003 &   68.3 \\
IRAS05373 &  84.100671  & -72.692390  &  84.100596  & -72.692388  & 0.10  &  645 &  0.8 &  643 &  0.6 & 0.54 & 0.01 &  9.736 &  0.004 &   44.1  & 0.50 & 0.01 &  8.162 &  0.003 &   31.5 \\
MSXLMC971 &  84.966000  & -70.021426  &  84.966000  & -70.021417  & 0.12  &  648 &  0.5 &  649 &  0.4 & 0.69 & 0.01 & 10.290 &  0.003 &   35.4  & 0.57 & 0.01 &  8.406 &  0.002 &   26.4 \\
MSXLMC937 &  85.150438  & -69.880528  &  85.150349  & -69.880501  & 0.19  &  691 &  0.8 &  691 &  0.8 & 0.62 & 0.01 &  8.890 &  0.003 &   67.3  & 0.58 & 0.01 &  7.478 &  0.003 &   67.9 \\
IRAS05416 &  85.336488  & -69.078812  &  85.336451  & -69.078794  & 0.18  &  901 &  0.9 &  907 &  0.9 & 1.07 & 0.01 & 12.479 &  0.004 &   48.8  & 0.92 & 0.01 &  9.451 &  0.003 &   97.7 \\
LMCLPV76711 &  85.571808  & -70.538970  &  85.571729  & -70.539004  & 0.17  &  552 &  0.5 &  550 &  0.4 & 0.49 & 0.02 &  9.274 &  0.003 &   21.7  & 0.43 & 0.01 &  8.024 &  0.002 &   16.4 \\
IRAS05495 &  87.250017  & -70.556247  &  87.249892  & -70.556277  & 0.43  & 11450 & 3220 & 6306 & 1753 & 0.05 & 0.02 & 15.472 &  0.010 &    2.2  & 0.03 & 0.02 & 13.434 &  0.010 &    2.3 \\
ERO0550261 &  87.608879  & -69.934185  &  87.608836  & -69.934213  & 0.31  & 1100 &  2.5 & 1115 &  1.3 & 0.60 & 0.02 & 15.518 &  0.010 &    3.2  & 0.69 & 0.01 & 11.527 &  0.004 &   32.3 \\
IRAS05515 &  87.708112  & -71.393213  &  87.708059  & -71.393231  & 0.13  &  664 &  0.6 &  667 &  0.4 & 0.62 & 0.01 & 10.281 &  0.004 &   45.8  & 0.54 & 0.01 &  8.504 &  0.002 &   28.9 \\
MSXLMC1797 &  87.889933  & -71.326194  &  87.889961  & -71.326161  & 0.17  &  686 &  2.2 &  704 &  1.4 & 0.74 & 0.03 & 15.589 &  0.017 &    9.0  & 0.42 & 0.02 & 11.548 &  0.007 &   74.7 \\
IRAS05568 &  89.161538  & -67.892749  &  89.161467  & -67.892763  & 0.18  & 1206 &  1.0 & 1211 &  0.9 & 1.27 & 0.01 & 12.090 &  0.003 &   65.0  & 1.06 & 0.01 &  9.069 &  0.002 &  100.9 \\
IRAS06028 &  90.688067  & -67.378653  &  90.688041  & -67.378667  & 0.12  &  860 &  0.4 &  860 &  0.3 & 0.84 & 0.01 &  9.755 &  0.002 &   58.8  & 0.71 & 0.01 &  7.723 &  0.001 &   57.7 \\
IRAS06108 &  92.544704  & -70.767581  &  92.544510  & -70.767579  & 0.25  &  527 &  0.3 &  528 &  0.3 & 0.54 & 0.01 & 10.179 &  0.003 &   34.2  & 0.46 & 0.01 &  8.512 &  0.002 &   31.1 \\
IRAS05506 &  87.485502  & -70.886597  &  87.485630  & -70.886601  & 0.17  & 1029 &  1.6 & 1018 &  1.3 & 0.77 & 0.02 &  9.416 &  0.006 &  141.5  & 0.70 & 0.02 &  7.563 &  0.005 &  129.8 \\
IRAS05125 &  78.003364  & -70.540098  &  78.003233  & -70.540035  & 0.28  & 1147 &  1.4 & 1147 &  1.4 & 0.67 & 0.01 & 11.073 &  0.005 &   54.5  & 0.55 & 0.01 &  8.522 &  0.004 &   58.4 \\
ERO0529379 &  82.407854  & -72.831365  &  82.407960  & -72.831345  & 0.22  &  685 &  1.4 &  686 &  1.7 & 0.54 & 0.02 & 14.179 &  0.010 &   20.7  & 0.50 & 0.03 & 10.793 &  0.009 &  131.5 \\
ERO0518117 &  79.548769  & -70.507508  &  79.548787  & -70.507469  & 0.32  & 9479 & 406.5 & 7935 & 143.5 & 0.96 & 0.04 & 14.598 &  0.055 &   15.0  & 0.99 & 0.03 & 11.471 &  0.024 &   16.2 \\
\hline
\end{tabular}
\tablefoot{
  Listed are the source name, the input coordinates used to search the WISE/NEOWISE archive, the "final" coordinates being the average of the coordinates
  of all entries contributing to the LC, the distance between input and final coordinates,
  periods in the W1 and W2 band, the amplitude, mean magnitude and reduced $\chi^2$ in the W1 band, the amplitude, mean magnitude
  and reduced $\chi^2$ in the W2 band. All error bars have been scaled to give a reduced $\chi^2$ of unity. \\
}\\

\vfill
\end{sidewaystable*}

  \clearpage

\section{SED fitting}
\label{App-SED}

\begin{multicols}{2}
Table~\ref{Tab-Phot} gives the references to the photometric data used to construct the SEDs.

\smallskip
Input to the MoD RT models are atmosphere models from \citet{Aringer09} 
with Z/Z$_{\odot}$= 0.10 and 0.33 for the SMC and LMC, respectively, $T_{\rm eff}$
between 2600 and 4000~K at 100~K intervals,
$\log g= $0.0~dex, and a fixed C/O ratio of 1.4 and mass of 2~\msol.
These abundances in the Aringer model grid are closest to the SMC and LMC abundances.
The photometry is de-reddened using the reddening map of \citet{Skowron21} and the $E(V-I)$ value in that map closest to the source is taken.
The visual extinction is then adopted as $A_{\rm V}= 3.1 \cdot E(V-I) /1.318$, following \citet{Skowron21}, and the reddening law of \citet{Cardelli89} is
used with $R_{\rm V}$= 3.1 at other wavelengths.

The dust around the C stars is assumed to be a combination of  AMC, SiC, and MgS.
The optical constants are taken from \citet{Zubko1996} (the ACAR species) 
for AMC,
\citealt{Pitman08} 
for SiC, and
\citet{Hofmeister03} 
for MgS.
The  absorption and scattering coefficients are calculated using a distribution of hollow spheres  (DHS, \citealt{Min05})
with assuming a typical grain size of 0.15~$\mu$m.
In DHS, the grains are modelled with a vacuum core which has a fractional volume uniformly distributed between zero and $f_{\rm max}$
in order to simulate the  fluffiness of real grains. \citet{GS18} found that $f_{\rm max} = 0.7$ fitted the data well, and we use that value here.
SiC and MgS are minor species relative to AMC, and the specific density of the dust grains is
around 1.5~g~cm$^{-3}$\footnote{$\rho_{\rm d}$ is 1.50~g~cm$^{-3}$ for a mix AMC:SiC:MgS=100:5:10 which was adopted in 27 objects, and the density
ranges from 1.30~g~cm$^{-3}$ for pure AMC (adopted in 2 sources) to 1.66~g~cm$^{-3}$, adopted in one source.}.

In the Nanni models the formation of AMC, SiC, and metallic iron dust species are considered.
The input optical properties and typical grain size for amorphous carbon dust have been selected to reproduce the colour-colour diagrams in the infrared and colour-magnitude diagrams
based on a combination of 2MASS and {\it Gaia} DR2 photometry (see \citealt{lebzelter2018new}; \citealt{Nanni2016, Nanni2019}).
The optical properties of AMC, SiC and metallic iron are from \citet{Jaeger98}, 
\citet{Pegourie1988} and \citet{Leksina67}, respectively.

Figure~\ref{AppFig-SED} shows examples of the fits to the SEDs. On the left side the MoD models and on the right side the Nanni models for a given star.
The complete set of figures is available via Zenodo.

\noindent
Figure~\ref{AppFig-Compare} compares MLRs, luminosities and GTD ratios between the two sets of RT models.
These models have been run independently and are run under slightly different assumptions.
For example, different optical properties of AMC and SiC have been adopted and different grain geometries (solid spheres versus a distribution of hollow spheres).
MgS is considered in MoD (but not metallic iron) and vice versa for the Nanni models.
The Nanni models fit only the observed photometry, while in the MoD models the mid-IR spectrum (when available) is also part of the fit
and used to manually adjust the fractions of SiC and MgS relative to AMC.
In MoD the temperature at the inner dust radius is allowed to vary freely, while in the Nanni models the dust formation and condensation process is followed
self-consistently which implies
typically values close to the expected condensation temperature.
In the Nanni models the luminosity is not fit freely, but the best match among a grid of luminosities is chosen.

Star symbols in Fig.~\ref{AppFig-Compare} indicate unreliable models and excludes stars with either poor fits to the SEDs (defined as the poorest 10\% in $\chi^2$ in
either the Nanni or MoD models), or very low dust temperatures at the inner radius (below 700~K in either the Nanni or MoD models), so that the hypothesis of a
dust driven wind is likely not directly applicable).
\end{multicols}

\begin{table*}[h]

  \small
  \centering
  
  \caption{\label{Tab-Phot} Photometry used to construct the SEDs.} 
    \begin{tabular}{llll}
  \hline  \hline
 Filters  & Instrument         &  Reference                     \\ 
\hline
 $V, I$                       & OGLE  & \citet{Soszynski09,Soszynski11} \\
 $B_{\rm p}$, $G$, $R_{\rm p}$ & {\it Gaia} & \citet{GaiaDR3Vallenari22} \\
 $Y, J, K$ & VMC DR6  & \citet{Cioni11} \\
 $J, H, K$ & 2MASS-6X &  \citet{Cutri12}\\
 $J, H, K$ & IRSF     &  \citet{Kato_IRSF}\\
 S7, S11, L15\tablefootmark{b} & Akari & \citet{Ita2010, Kato12}   \\ 
 9, 18~$\mu$m & Akari & \cite{AKIRC10} \\
 W1, W2, W3, W4\tablefootmark{c} & AllWISE & \citet{Cutri_Allwise} \\
 3.6, 4.5, 5.8, 8.5~$\mu$m & {\it Spitzer} IRAC & IPAC\tablefootmark{d}, \citet{Gruendl08,Whitney08} \\
 A,C,D             &  MSX & \citet{Egan03} & \\
 12, 25~$\mu$m    & IRAS PSC \& FSC & \citet{Beichmann85,MoshirFSC} \\ 
 24, 70~$\mu$m    & {\it Spitzer} MIPS         & IRSA\tablefootmark{e}, \citet{Gruendl08}   \\
 70 -- 350~$\mu$m & {\it Herschel} PACS, SPIRE & \citet{Seale14} \\
 1330~$\mu$m      & ALMA & This work  \\
 4.5, 6.7, 12~$\mu$m & ISOCAM  & \citet{Tanabe04}, for NGC419 MIR1 \\
 \hline
    \end{tabular}
\tablefoot{
\tablefoottext{a}{For the S7, S11, and L15 filters only errors in the magnitudes of $<$0.15, $<$0.20, and $<$0.20~mag were accepted, respectively;}
\tablefoottext{b}{In the W3 and W4 filters only errors in the magnitudes of $<$0.30, and $<$0.25~mag were accepted, respectively;}
\tablefoottext{c}{VizieR catalog II/305/catalog;}
\tablefoottext{d}{\url{https://irsa.ipac.caltech.edu/applications/Gator/}, the "SAGE MIPS 24 um Epoch 1 and Epoch 2 Full List".}
}
\end{table*}

\FloatBarrier
\clearpage

\begin{figure*}
  \centering
 
\begin{minipage}{0.38\textwidth}
\resizebox{\hsize}{!}{\includegraphics[angle=-0]{MSXLMC527_sed.ps}} 
\end{minipage}
\begin{minipage}{0.39\textwidth}
\resizebox{\hsize}{!}{\includegraphics[angle=-0]{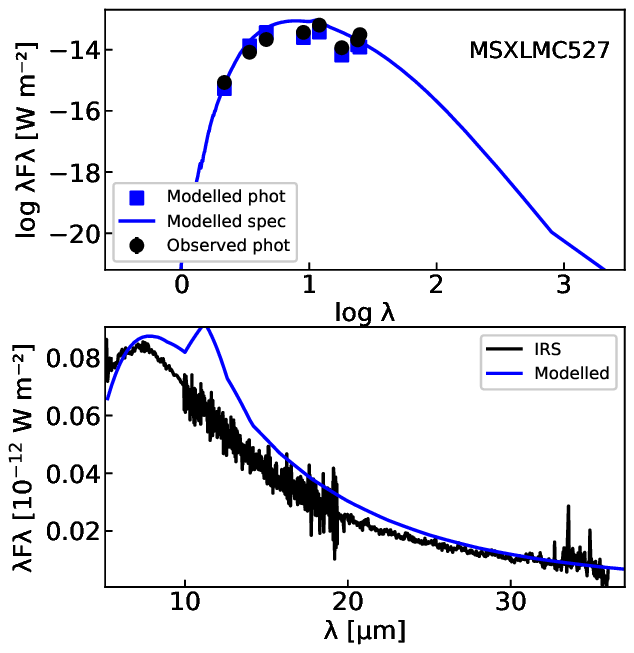}} 
\end{minipage}

\begin{minipage}{0.38\textwidth}
\resizebox{\hsize}{!}{\includegraphics[angle=-0]{IRAS06108_sed.ps}} 
\end{minipage}
\begin{minipage}{0.39\textwidth}
\resizebox{\hsize}{!}{\includegraphics[angle=-0]{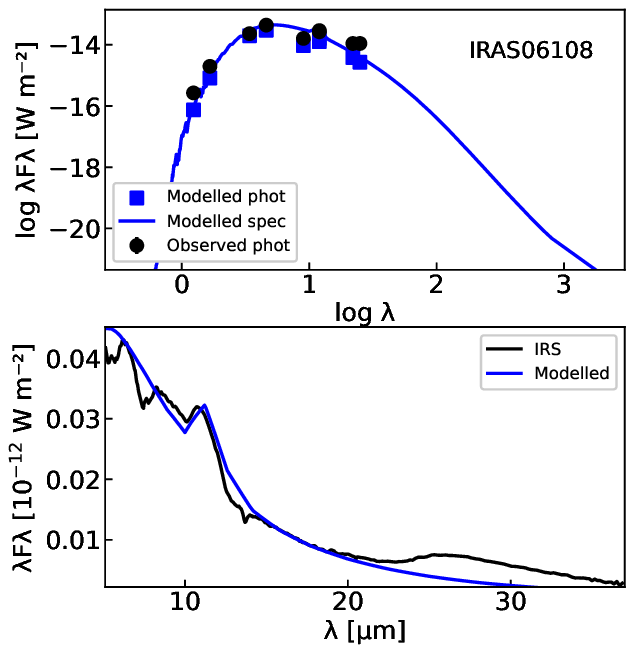}} 
\end{minipage}

\begin{minipage}{0.38\textwidth}
\resizebox{\hsize}{!}{\includegraphics[angle=-0]{IRAS05568_sed.ps}} 
\end{minipage}
\begin{minipage}{0.39\textwidth}
\resizebox{\hsize}{!}{\includegraphics[angle=-0]{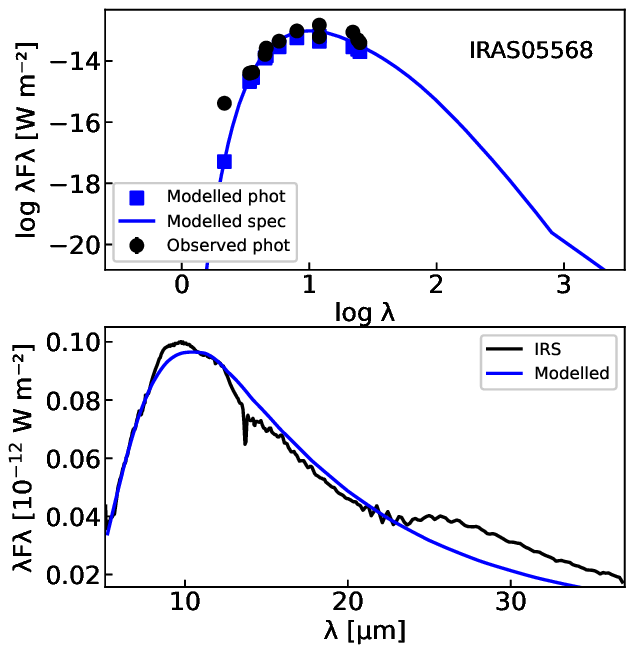}} 
\end{minipage}

\caption{\label{AppFig-SED} Examples of SED fits.
  On the left the MoD models and on the right the Nanni models.
  In the MoD models the model spectra are scaled to match the observed spectra based on the average flux
  in the 6.35-6.55~$\mu$m region, in the Nanni models the spectra are not included in the fitting, but the scaling is applied.
  The complete set of figures is available via Zenodo.}
\end{figure*}

\setcounter{figure}{0}

\begin{figure*}
  \centering
  
\begin{minipage}{0.38\textwidth}
\resizebox{\hsize}{!}{\includegraphics[angle=-0]{NGC419MIR1_sed.ps}} 
\end{minipage}
\begin{minipage}{0.39\textwidth}
\resizebox{\hsize}{!}{\includegraphics[angle=-0]{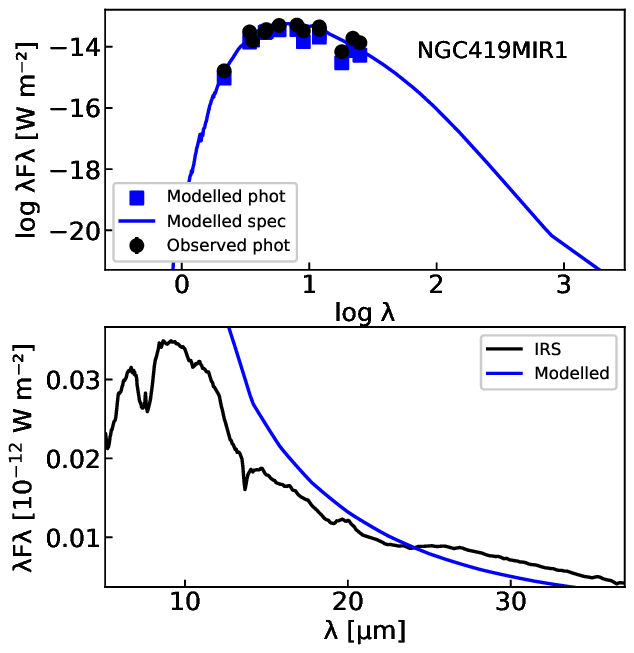}} 
\end{minipage}

\begin{minipage}{0.37\textwidth}
\resizebox{\hsize}{!}{\includegraphics[angle=-0]{IRAS05133_sed.ps}} 
\end{minipage}
\begin{minipage}{0.38\textwidth}
\resizebox{\hsize}{!}{\includegraphics[angle=-0]{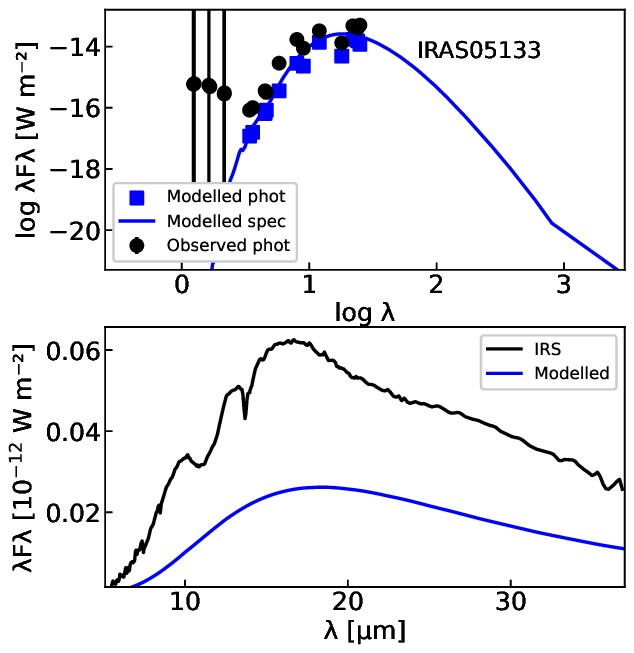}} 
\end{minipage}

\caption{continued.}
\end{figure*}

\begin{figure*}
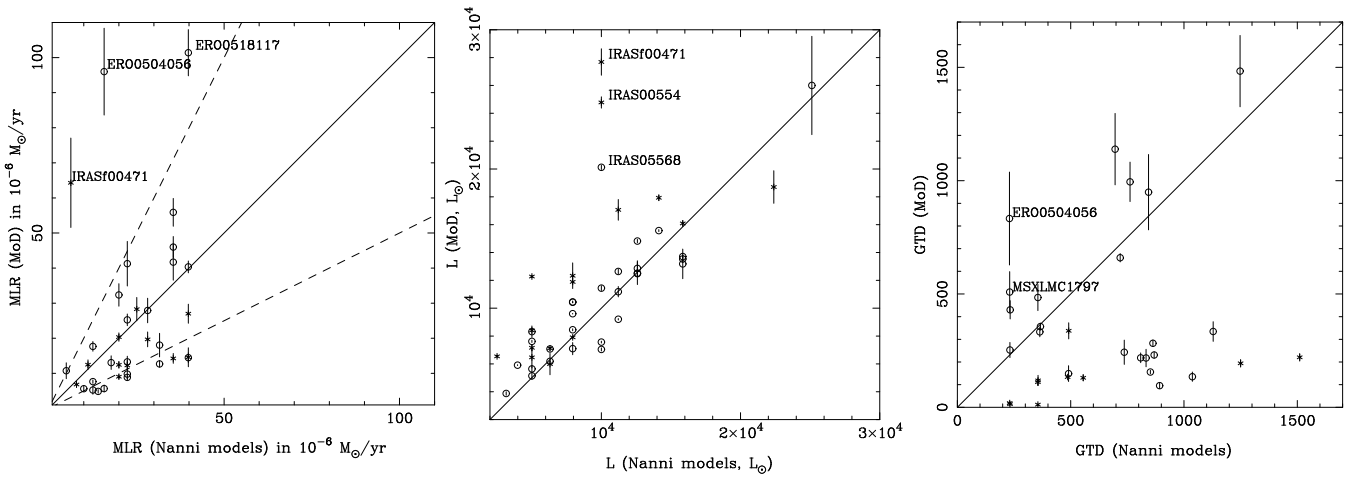

  \centering
  
\begin{minipage}{0.31\textwidth}
\resizebox{\hsize}{!}{\includegraphics[angle=-0]{MLR_MLR.ps}} 
\end{minipage}
\begin{minipage}{0.33\textwidth}
\resizebox{\hsize}{!}{\includegraphics[angle=-0]{Lum_Lum.ps}} 
\end{minipage}
\begin{minipage}{0.31\textwidth}
\resizebox{\hsize}{!}{\includegraphics[angle=-0]{GTD_GTD.ps}} 
\end{minipage}

\caption{\label{AppFig-Compare} Comparison of the fitting results from MoD and the Nanni models for the MLR (left),
  luminosity (middle), and GTD ratio (right-hand panel).
  The one-to-one line is drawn in all panels, and in the two left-most panels lines that indicate factors of 0.5 and 2 difference as well.
Outliers are labelled. Star symbols indicate unreliable models (see text in Appendix~\ref{App-SED}). }
\end{figure*}

\FloatBarrier
\clearpage

\begin{sidewaystable*}[h]
\section{Additional tables}

  \fontsize{6.8}{8.2}\selectfont

  \setlength{\tabcolsep}{1.0mm}
\caption{\label{Tab-final} Results of the ALMA observations.}
\begin{tabular}{llccccccccccccccrc}
\hline
\hline
\centering
Name & SG &  RA    & Dec   &  Source size       & Continuum  & Area      & $V_{\star}$  & $\Delta V$   & Peak   & RMS      & Area      & RMS     \\
     &   &  (h:m:s)  & (d:m:s) & (\arcsec)          &  (mJy)     & (Jy \kms) & (\kms)       & (\kms)        & (Jy)   & (Jy)     & (Jy km/s)  &  (Jy)           \\
\hline
IRASf00471 &  SMC  &  0:48:59.491 $\pm$ 0.024  &  -73:35:38.832 $\pm$ 0.035  &  0.475 $\pm$ 0.083 x 0.398 $\pm$ 0.056  & $<$ 0.039    &  0.169 $\pm$ 0.032  &  171.66 $\pm$ 0.65  &  13.40 $\pm$ 1.86  &  0.019 $\pm$ 0.003  &  0.007  &  $<$ 0.093  &  0.007   \\
IRAS00554  &  SMC  &  \tablefootmark{a}    &                 &    & $<$ 0.036    &  $<$ 0.115  &  170.00 f  &  18.00 f  &  $<$ 0.010  &  0.007  &  $<$ 0.096  &  0.007   \\
NGC419MIR1 &  SMC  &  1:08:17.479 $\pm$ 0.023  &  -72:53:09.413 $\pm$ 0.023  &  0.658 $\pm$ 0.059 x 0.587 $\pm$ 0.048  & $<$ 0.048    &  0.586 $\pm$ 0.033  &  174.08 $\pm$ 0.32  &  19.82 $\pm$ 0.82  &  0.044 $\pm$ 0.003  &  0.007  &  $<$ 0.096  &  0.006   \\
IRAS04340 &  LMCwk  &  4:33:43.766 $\pm$ 0.179  &  -70:09:50.573 $\pm$ 0.066  &  1.168 $\pm$ 0.422 x 0.672 $\pm$ 0.154  & $<$ 0.027    &  0.147 $\pm$ 0.033  &  251.49 $\pm$ 0.98  &  21.39 $\pm$ 2.95  &  0.010 $\pm$ 0.002  &  0.005  &  $<$ 0.060  &  0.004   \\
IRAS04375 &  LMCst  &  4:36:44.503 $\pm$ 0.035  &  -72:42:01.051 $\pm$ 0.046  &  0.804 $\pm$ 0.111 x 0.664 $\pm$ 0.079  &  0.122  $\pm$  0.027  &  1.073 $\pm$ 0.117  &  244.27 $\pm$ 1.52  &  44.54 $\pm$ 4.90  &  0.036 $\pm$ 0.004  &  0.015  &  $<$ 0.315  &  0.015   \\
MSXLMC1119 &  LMCwk  &  4:47:38.848 $\pm$ 0.050  &  -69:21:16.981 $\pm$ 0.060  &  1.066 $\pm$ 0.163 x 0.755 $\pm$ 0.087  &  0.077  $\pm$  0.011  &  0.559 $\pm$ 0.038  &  253.69 $\pm$ 0.74  &  37.52 $\pm$ 1.84  &  0.022 $\pm$ 0.002  &  0.005  &  $<$ 0.135  &  0.006   \\
IRAS04496 &  LMCwk  &  \tablefootmark{a}    &                 &    & $<$ 0.039    &  $<$ 0.029  &  250.00 f  &  30.00 f  &  $<$ 0.001  &  0.005  &  $<$ 0.084  &  0.005   \\
IRAS04523 &  LMCwk  &  4:51:50.919 $\pm$ 0.017  &  -70:38:32.064 $\pm$ 0.019  &  0.937 $\pm$ 0.051 x 0.752 $\pm$ 0.034  & $<$ 0.045    &  1.513 $\pm$ 0.038  &  221.64 $\pm$ 0.26  &  35.86 $\pm$ 0.80  &  0.063 $\pm$ 0.002  &  0.008  &  $<$ 0.141  &  0.005   \\
SAGEMCJ045344 &  LMCst  &  \tablefootmark{a}  &                             &                   & $<$ 0.195    &  $<$ 0.504  &  250.00 f  &  30.00 f  &  $<$ 0.025  &  0.016  &  $<$ 0.258  &  0.014   \\
IRAS04557 &  LMCst  &  4:55:38.981 $\pm$ 0.026  &  -67:49:10.528 $\pm$ 0.015  &  0.628 $\pm$ 0.062 x 0.469 $\pm$ 0.034  & $<$ 0.072    &  1.413 $\pm$ 0.132  &  293.28 $\pm$ 1.15  &  38.90 $\pm$ 3.32  &  0.054 $\pm$ 0.004  &  0.016  &  $<$ 0.432  &  0.017   \\
MSXLMC1220 &  LMCwk  &  \tablefootmark{a}    &                 &  & $<$ 0.033    &  $<$ 0.143  &  250.00 f  &  30.00 f  &  $<$ 0.007  &  0.006  &  $<$ 0.114  &  0.006   \\
MSXLMC1303 &  LMCwk  &  4:58:42.042 $\pm$ 0.166  &  -68:07:15.006 $\pm$ 0.049  &  1.145 $\pm$ 0.399 x 0.522 $\pm$ 0.086  &  0.093  $\pm$  0.012  &  0.275 $\pm$ 0.021  &  248.10 $\pm$ 0.33  &  15.59 $\pm$ 0.94  &  0.026 $\pm$ 0.002  &  0.006  &  $<$ 0.063  &  0.005   \\
MSXLMC1282 &  LMCwk  &  5:01:00.900 $\pm$ 0.038  &  -67:35:23.676 $\pm$ 0.053  &  1.162 $\pm$ 0.143 x 0.686 $\pm$ 0.055  & $<$ 0.030    &  0.667 $\pm$ 0.041  &  241.21 $\pm$ 1.26  &  69.67 $\pm$ 3.00  &  0.014 $\pm$ 0.001  &  0.004  &  $<$ 0.138  &  0.005   \\
ERO0502315 &  LMCst  &  5:02:31.504 $\pm$ 0.202  &  -68:05:35.880 $\pm$ 1.050  &  2.387 $\pm$ 2.474 x 0.691 $\pm$ 0.469  & $<$ 0.078    &  1.164 $\pm$ 0.094  &  218.59 $\pm$ 0.49  &  21.56 $\pm$ 1.40  &  0.080 $\pm$ 0.006  &  0.016  &  $<$ 0.249  &  0.017   \\
MSXLMC91 &  LMCst  &  5:03:38.573 $\pm$ 0.055  &  -68:53:12.710 $\pm$ 0.057  &  0.701 $\pm$ 0.135 x 0.682 $\pm$ 0.128  & $<$ 0.072    &  2.200 $\pm$ 0.122  &  258.30 $\pm$ 0.82  &  44.20 $\pm$ 2.63  &  0.075 $\pm$ 0.004  &  0.015  &  $<$ 0.451  &  0.017   \\
TRM74    &  LMCwk  &  5:03:43.012 $\pm$ 0.017  &  -66:44:56.530 $\pm$ 0.019  &  0.916 $\pm$ 0.047 x 0.807 $\pm$ 0.037  & $<$ 0.030    &  1.442 $\pm$ 0.048  &  280.32 $\pm$ 0.26  &  25.15 $\pm$ 1.34  &  0.086 $\pm$ 0.003  &  0.009  &  $<$ 0.084  &  0.005   \\
ERO0504056 &  LMCst  &  5:04:05.621 $\pm$ 0.037  &  -68:23:40.312 $\pm$ 0.030  &  0.601 $\pm$ 0.086 x 0.536 $\pm$ 0.069  & $<$ 0.066    &  0.656 $\pm$ 0.089  &  261.48 $\pm$ 0.59  &  14.88 $\pm$ 1.76  &  0.066 $\pm$ 0.008  &  0.017  &  $<$ 0.237  &  0.016   \\
MSXLMC87 &  LMCwk  &  5:10:19.656 $\pm$ 0.053  &  -69:49:51.506 $\pm$ 0.075  &  1.042 $\pm$ 0.179 x 0.837 $\pm$ 0.121  &  0.080  $\pm$  0.008  &  0.696 $\pm$ 0.051  &  230.56 $\pm$ 1.62  &  69.20 $\pm$ 4.27  &  0.015 $\pm$ 0.001  &  0.005  &  $<$ 0.195  &  0.005   \\
IRAS05113 &  LMCwk  &  5:10:56.854 $\pm$ 0.033  &  -69:35:30.512 $\pm$ 0.049  &  0.952 $\pm$ 0.118 x 0.747 $\pm$ 0.075  & $<$ 0.030    &  0.522 $\pm$ 0.045  &  217.20 $\pm$ 2.47  &  66.98 $\pm$ 7.39  &  0.012 $\pm$ 0.001  &  0.005  &  $<$ 0.152  &  0.005   \\
IRAS05132 &  LMCwk  &  5:12:51.062 $\pm$ 0.025  &  -69:37:50.351 $\pm$ 0.032  &  0.970 $\pm$ 0.084 x 0.714 $\pm$ 0.047  &  0.035  $\pm$  0.008  &  0.874 $\pm$ 0.038  &  255.15 $\pm$ 0.39  &  39.62 $\pm$ 1.11  &  0.033 $\pm$ 0.001  &  0.008  &  $<$ 0.111  &  0.005   \\
IRAS05133 &  LMCst  &  5:13:01.786 $\pm$ 0.025  &  -69:33:50.951 $\pm$ 0.025  &  0.605 $\pm$ 0.067 x 0.518 $\pm$ 0.049  &  0.088  $\pm$  0.024  &  1.071 $\pm$ 0.096  &  227.27 $\pm$ 0.75  &  26.12 $\pm$ 3.34  &  0.061 $\pm$ 0.005  &  0.014  &  $<$ 0.292  &  0.015   \\
ERO0518484 &  LMCst  &  5:18:48.404 $\pm$ 0.031  &  -69:33:34.650 $\pm$ 0.041  &  0.577 $\pm$ 0.098 x 0.497 $\pm$ 0.072  & $<$ 0.078    &  0.412 $\pm$ 0.087  &  237.98 $\pm$ 0.60  &  16.05 $\pm$ 1.50  &  0.037 $\pm$ 0.008  &  0.014  &  $<$ 0.213  &  0.017   \\
IRAS05190 &  LMCwk  &  5:18:56.271 $\pm$ 0.010  &  -67:45:04.475 $\pm$ 0.010  &  0.854 $\pm$ 0.026 x 0.754 $\pm$ 0.021  & $<$ 0.030    &  2.354 $\pm$ 0.043  &  236.70 $\pm$ 0.23  &  37.84 $\pm$ 0.86  &  0.093 $\pm$ 0.002  &  0.008  &  $<$ 0.123  &  0.006   \\
TRM88 &  LMCwk  &  5:20:19.410 $\pm$ 0.066  &  -66:35:47.854 $\pm$ 0.039  &  0.864 $\pm$ 0.159 x 0.639 $\pm$ 0.087  & $<$ 0.036    &  0.344 $\pm$ 0.044  &  260.55 $\pm$ 2.13  &  62.05 $\pm$ 7.51  &  0.008 $\pm$ 0.001  &  0.005  &  $<$ 0.144  &  0.006   \\
MSXLMC527 &  LMCwk  &  \tablefootmark{a}  &                               &                       & $<$ 0.033    &  $<$ 0.176  &  290.00 f  &  30.00 f  &  $<$ 0.009  &  0.007  &  $<$ 0.110  &  0.005   \\
ERO0525406 &  LMCst  &  5:25:40.655 $\pm$ 0.035  &  -70:08:27.244 $\pm$ 0.040  &  0.594 $\pm$ 0.111 x 0.424 $\pm$ 0.055  & $<$ 0.102    &  0.498 $\pm$ 0.091  &  227.86 $\pm$ 1.06  &  17.41 $\pm$ 2.94  &  0.041 $\pm$ 0.010  &  0.015  &  $<$ 0.300  &  0.016   \\
MSXLMC474 &  LMCst  &  \tablefootmark{a}  &                               &    & $<$ 0.069    &  $<$ 0.399  &  250.00 f  &  30.00 f  &  $<$ 0.020  &  0.015  &  $<$ 0.375  &  0.017   \\
IRAS05315 &  LMCst  &  5:30:44.134 $\pm$ 0.019  &  -71:43:00.509 $\pm$ 0.019  &  0.612 $\pm$ 0.046 x 0.607 $\pm$ 0.045  & $<$ 0.072    &  1.594 $\pm$ 0.088  &  239.74 $\pm$ 0.43  &  25.03 $\pm$ 1.26  &  0.096 $\pm$ 0.006  &  0.015  &  $<$ 0.243  &  0.013   \\
MSXLMC1780 &  LMCwk  &  5:33:01.796 $\pm$ 0.033  &  -68:23:58.434 $\pm$ 0.038  &  1.097 $\pm$ 0.089 x 1.011 $\pm$ 0.078  &  0.055  $\pm$  0.011  &  1.653 $\pm$ 0.143  &  256.76 $\pm$ 1.00  &  31.70 $\pm$ 3.08  &  0.078 $\pm$ 0.004  &  0.015  &  $<$ 0.189  &  0.006   \\
IRAS05373 &  LMCst  &  5:36:24.161 $\pm$ 0.026  &  -72:41:32.604 $\pm$ 0.037  &  0.704 $\pm$ 0.089 x 0.572 $\pm$ 0.060  & $<$ 0.099    &  1.078 $\pm$ 0.106  &  210.73 $\pm$ 0.73  &  27.25 $\pm$ 2.24  &  0.059 $\pm$ 0.005  &  0.016  &  $<$ 0.222  &  0.014   \\
MSXLMC971 &  LMCwk  &  5:39:51.840 $\pm$ 0.035  &  -70:01:17.134 $\pm$ 0.037  &  0.992 $\pm$ 0.098 x 0.820 $\pm$ 0.069  & $<$ 0.027    &  0.910 $\pm$ 0.034  &  254.24 $\pm$ 0.26  &  28.79 $\pm$ 0.66  &  0.047 $\pm$ 0.002  &  0.006  &  $<$ 0.115  &  0.005   \\
MSXLMC937 &  LMCwk  &  5:40:36.105 $\pm$ 0.038  &  -69:52:49.901 $\pm$ 0.049  &  1.095 $\pm$ 0.130 x 0.765 $\pm$ 0.068  & $<$ 0.021    &  0.854 $\pm$ 0.055  &  260.60 $\pm$ 0.46  &  50.35 $\pm$ 1.98  &  0.025 $\pm$ 0.002  &  0.006  &  $<$ 0.120  &  0.006   \\
IRAS05416 &  LMCwk  &  5:41:20.757 $\pm$ 0.015  &  -69:04:43.723 $\pm$ 0.024  &  0.873 $\pm$ 0.057 x 0.679 $\pm$ 0.035  & $<$ 0.039    &  1.300 $\pm$ 0.033  &  236.04 $\pm$ 0.25  &  27.85 $\pm$ 0.87  &  0.070 $\pm$ 0.002  &  0.005  &  $<$ 0.084  &  0.005   \\
LMCLPV76711 &  LMCwk  &  5:42:17.234 $\pm$ 0.099  &  -70:32:20.292 $\pm$ 0.099  &  1.121 $\pm$ 0.295 x 0.758 $\pm$ 0.148  & $<$ 0.030    &  0.253 $\pm$ 0.048  &  255.54 $\pm$ 1.62  &  26.58 $\pm$ 5.62  &  0.014 $\pm$ 0.002  &  0.006  &  $<$ 0.099  &  0.006   \\
IRAS05495 &  LMCst  &  5:49:00.004 $\pm$ 0.014  &  -70:33:22.489 $\pm$ 0.017  &  0.657 $\pm$ 0.040 x 0.595 $\pm$ 0.033  &  0.116  $\pm$  0.019  &  2.275 $\pm$ 0.109  &  246.85 $\pm$ 0.42  &  31.43 $\pm$ 1.23  &  0.109 $\pm$ 0.005  &  0.016  &  $<$ 0.336  &  0.016   \\
ERO0550261 &  LMCst  &  5:50:26.131 $\pm$ 0.019  &  -69:56:03.066 $\pm$ 0.022  &  0.723 $\pm$ 0.056 x 0.580 $\pm$ 0.037  & $<$ 0.111    &  1.828 $\pm$ 0.104  &  270.39 $\pm$ 0.67  &  31.39 $\pm$ 2.05  &  0.087 $\pm$ 0.005  &  0.015  &  $<$ 0.300  &  0.015   \\
IRAS05515 &  LMCst  &  5:50:49.947 $\pm$ 0.034  &  -71:23:35.567 $\pm$ 0.041  &  0.770 $\pm$ 0.102 x 0.645 $\pm$ 0.074  & $<$ 0.099    &  1.536 $\pm$ 0.091  &  243.00 $\pm$ 0.77  &  41.52 $\pm$ 2.07  &  0.056 $\pm$ 0.004  &  0.013  &  $<$ 0.309  &  0.015   \\
MSXLMC1797 &  LMCwk  &  5:51:33.584 $\pm$ 0.098  &  -71:19:34.298 $\pm$ 0.119  &  1.260 $\pm$ 0.317 x 0.884 $\pm$ 0.177  &  0.114  $\pm$  0.012  &  0.550 $\pm$ 0.034  &  249.45 $\pm$ 0.33  &  19.04 $\pm$ 1.42  &  0.043 $\pm$ 0.002  &  0.005  &  $<$ 0.084  &  0.006   \\
IRAS05568 &  LMCwk  &  5:56:38.769 $\pm$ 0.015  &  -67:53:33.896 $\pm$ 0.023  &  0.949 $\pm$ 0.055 x 0.738 $\pm$ 0.034  & $<$ 0.042    &  1.498 $\pm$ 0.033  &  267.28 $\pm$ 0.19  &  28.20 $\pm$ 0.56  &  0.080 $\pm$ 0.002  &  0.008  &  \tablefootmark{b}  &  0.006   \\
IRAS06028 &  LMCst  &  6:02:45.136 $\pm$ 0.020  &  -67:22:43.151 $\pm$ 0.020  &  0.610 $\pm$ 0.048 x 0.604 $\pm$ 0.047  & $<$ 0.111    &  1.898 $\pm$ 0.122  &  281.52 $\pm$ 0.71  &  40.58 $\pm$ 1.90  &  0.070 $\pm$ 0.004  &  0.015  &  $<$ 0.354  &  0.017   \\
IRAS06108 &  LMCwk  &  6:10:10.729 $\pm$ 0.064  &  -70:46:03.292 $\pm$ 0.066  &  0.937 $\pm$ 0.171 x 0.816 $\pm$ 0.133  & $<$ 0.030    &  0.476 $\pm$ 0.031  &  256.93 $\pm$ 0.34  &  18.76 $\pm$ 0.95  &  0.038 $\pm$ 0.002  &  0.006  &  $<$ 0.090  &  0.006   \\
\hline
\end{tabular}
\tablefoot{
  Listed are the source name, the SG (see Tab.~\ref{Tab-SG}), the fitted coordinates with errors from the ALMA image (Fig.~\ref{App-Fig-CO}),
  the major and minor axis of the source size (not deconvolved with the beam),
  the continuum flux at 1330~$\mu$m and error, the integrated area of the $^{12}$CO emission with error, the stellar velocity with error,
  the full width at zero intensity of the $^{12}$CO profile with error, the peak intensity of the profile with error, the rms noise level,
  the  upper limit to the integrated area of the $^{13}$CO emission, and the rms level. \\
}\\
\tablefoottext{a}{  \fontsize{9}{10.8}\selectfont
Input coordinates (Ra, Dec) of the objects that were not detected:
  IRAS 00554    (0:57:03.928, $-$73:35:14.593),
  IRAS 04496    (4:49:18.460, $-$69:53:14.494), 
  SAGEMCJ045344 (4:53:44.280, $-$66:11:45.953),  
  MSXLMC1220    (4:55:41.779, $-$68:57:22.621), 
  MSXLMC527     (5:22:19.740, $-$65:43:18.898), 
  MSXLMC474     (5:25:51.840, $-$68:46:34.207), 
}\\
\tablefoottext{b}{ \fontsize{9}{10.8}\selectfont
  IRAS 05568 was detected in $^{13}$CO with an integrated intensity 0.81 $\pm$ 0.04 \jks, $V_{\star}$ = 267.33 $\pm$ 0.35~\ks,
  and a peak of 0.043 $\pm$ 0.002~Jy  (for a fixed $V_{\rm exp}$ of 14.1~\ks).}
\vfill
\end{sidewaystable*}

\begin{sidewaystable*}

  \small
 \setlength{\tabcolsep}{1.3mm}

\caption{\label{Tab-SED} Results of the SED modelling and adopted pulsation periods.}
\begin{tabular}{lrrrrrrrrrrrrrrr}
\hline \hline
Name            & $L$     &  GTD    & \mdot $\times 10^{6}$ &  $V$  & $T_{\rm c}$ &  $L$     &  GTD & \mdot $\times 10^{6}$  & $T_{\rm c}$ & $L$      &  GTD    & \mdot $\times 10^{6}$ & \mdot $\times 10^{6}$   & Period \\  
                & (\lsol) &         & (\msolyr)             & (\ks) &  (K)       & (\lsol) &      & (\msolyr)              & (K)        &  (\lsol) &         & (\msolyr)             & (\msolyr)              & (d)    \\
\hline 
               & \multicolumn{5}{c}{Nanni models} & \multicolumn{4}{c}{MoD} &  \multicolumn{3}{c}{Adopted} & \multicolumn{1}{c}{CO model} &          \\
                \cmidrule(lr){2-6} \cmidrule(lr){7-10} \cmidrule(lr){11-13} \cmidrule(lr){14-14}  \\
     IRASf00471 & 10000 & 1889 &  6.3 &  8.1 & 1164 & 27685 $\pm$   950 &  9690 $\pm$  2973 &  64.3 $\pm$ 12.7 & 1107 $\pm$   17 & 16639 & 4279 &   20 &      9.6  &  680.0  $\pm$  0.6  \\ 
      IRAS00554 & 10000 & 1109 & 11.2 & 15.0 & 1051 & 24779 $\pm$   405 &  2072 $\pm$  1373 &  43.5 $\pm$ 15.1 &  865 $\pm$   10 & 15741 & 1516 &   22 & $<$  8.0  &  893.2  $\pm$  0.7  \\ 
     NGC419MIR1 & 10000 &  762 & 22.4 & 10.8 & 1013 &  7566 $\pm$ \phantom{2}68 &  995 $\pm$ \phantom{22}87 & 25.3 $\pm$ 1.7 & 1000 F &  8698 &  871 & 24 &  32.0  &  717.8  $\pm$  0.5  \\  
      IRAS04340 &  5012 & 1670 &  5.0 &  7.3 & 1164 &  8314 $\pm$   411 &  2409 $\pm$ \phantom{2}756 &  10.7 $\pm$  2.3 & 1116 $\pm$   24 &  6455 & 2006 &    7 &      5.1  &  439.6  $\pm$  0.3  \\ 
      IRAS04375 & 10000 &  491 & 15.8 & 21.7 &  999 &  7040 $\pm$   112 &   149 $\pm$    35 &   5.7 $\pm$  1.0 &  805 $\pm$   14 &  8391 &  270 &    9 &     36.5  &  693.8  $\pm$  0.4  \\ 
     MSXLMC1119 &  7943 & 1129 & 10.0 & 17.1 & 1061 &  7095 $\pm$   437 &   334 $\pm$    43 &   5.6 $\pm$  0.8 &  906 $\pm$   43 &  7507 &  614 &    7 &     19.8  &  544.4  $\pm$  0.2  \\ 
      IRAS04496 & 25120 & 1885 & 15.8 & 26.0 & 1166 & 25999 $\pm$  3529 &   686 $\pm$   853 &  18.7 $\pm$ 13.2 &  708 $\pm$   44 & 25556 & 1137 &   17 & $<$  1.6  &                     \\ 
      IRAS04523 & 22390 & 1511 & 35.5 & 18.6 & 1046 & 18705 $\pm$  1171 &   221 $\pm$    17 &  14.3 $\pm$  1.4 &  665 $\pm$   29 & 20465 &  577 &   23 &     43.8  &  874.4  $\pm$  0.5  \\ 
  SAGEMCJ045344 &  7943 &  973 & 17.8 &  4.0 & 1001 &  8451 $\pm$    74 &   249 $\pm$   285 &  14.9 $\pm$  7.9 &  800 $\pm$   14 &  8193 &  493 &   16 & $<$ 16.4  &  761.0  $\pm$  0.3  \\ 
      IRAS04557 & 12590 &  833 & 22.4 & 14.8 & 1003 & 14824 $\pm$   121 &   217 $\pm$    39 &  13.3 $\pm$  1.6 &  746 $\pm$ \phantom{1}9 & 13662 &  426 &   17 &     43.0  &  729.7  $\pm$  0.3  \\ 
     MSXLMC1220 &  5012 & 1629 &  3.2 &  4.0 & 1244 &  6471 $\pm$   761 &   477 $\pm$   586 &   5.6 $\pm$  3.8 &  862 $\pm$   90 &  5695 &  881 &    4 & $<$  5.8  &  513.7  $\pm$  0.3  \\ 
     MSXLMC1303 & 11220 &  696 & 35.5 &  8.9 &  999 & 11186 $\pm$   373 &  1139 $\pm$   158 &  41.7 $\pm$  5.1 &  838 $\pm$   21 & 11203 &  890 &   38 &      7.4  & 1005.0  $\pm$  3.1  \\ 
     MSXLMC1282 & 11220 &  555 &  7.9 & 35.1 & 1052 & 17063 $\pm$   745 &   130 $\pm$    14 &   6.8 $\pm$  0.8 &  851 $\pm$   13 & 13837 &  269 &    7 &     30.4  &  608.5  $\pm$  0.5  \\ 
     ERO0502315 &  7943 &  355 & 39.8 &  9.8 &  892 &  7910 $\pm$    65 &   111 $\pm$    15 &  27.0 $\pm$  2.7 &  554 $\pm$   10 &  7926 &  199 &   33 &     28.0  &                \\ 
       MSXLMC91 & 12590 & 1037 & 22.4 & 21.3 & 1033 & 12516 $\pm$   817 &   135 $\pm$    21 &   9.8 $\pm$  1.6 &  705 $\pm$   35 & 12553 &  374 &   15 &     65.6  &  897.0  $\pm$  0.5  \\ 
          TRM74 &  5012 &  231 & 22.4 & 10.9 &  893 &  7160 $\pm$   228 &    19 $\pm$     2 &  12.3 $\pm$  1.4 &  318 $\pm$   12 &  5990 &   65 &   17 &     35.8  &                \\ 
     ERO0504056 &  3981 &  230 & 15.8 &  5.9 &  898 &  5911 $\pm$    31 &   833 $\pm$   205 &  96.0 $\pm$ 12.4 & 1000 F        &  4851 &  438 &   39 &     14.7  &                \\ 
       MSXLMC87 & 15850 &  892 & 14.1 & 34.0 & 1133 & 13536 $\pm$   270 &    96 $\pm$    13 &   4.9 $\pm$  0.5 &  764 $\pm$   12 & 14647 &  292 &    8 &     31.4  &  615.4  $\pm$  1.2  \\ 
      IRAS05113 & 15850 &  555 & 12.6 & 39.8 & 1044 & 13183 $\pm$  1062 &    81 $\pm$    22 &   5.3 $\pm$  1.2 &  733 $\pm$   43 & 14455 &  212 &    8 &     24.4  &  630.6  $\pm$  0.7  \\ 
      IRAS05132 & 11220 &  868 & 22.4 & 18.5 & 1011 &  9207 $\pm$    54 &   231 $\pm$    14 &   8.8 $\pm$  0.5 &  835 $\pm$ \phantom{1}8 & 10164 &  448 &   14 &     29.3  &  681.7  $\pm$  0.4  \\ 
      IRAS05133 &  2512 &  118 & 39.8 &  8.3 &  363 &  6544 $\pm$    89 &    42 $\pm$    11 &  14.6 $\pm$  2.7 &  449 $\pm$   12 &  4054 &   70 &   24 &     28.5  &                \\ 
     ERO0518484 &  6310 &  356 & 28.2 &  5.8 &  908 &  7118 $\pm$    14 &   119 $\pm$    23 &  19.7 $\pm$  2.1 &  430 $\pm$ \phantom{1}3 &  6702 &  205 &   24 &     10.4  &  803.0  $\pm$ 22.3  \\ 
      IRAS05190 & 12590 &  852 & 31.6 & 19.5 & 1001 & 12482 $\pm$   130 &   156 $\pm$     8 &  12.7 $\pm$  0.9 &  705 $\pm$   12 & 12536 &  364 &   20 &     64.6  &  953.1  $\pm$  0.9  \\ 
          TRM88 &  7943 &  490 &  7.9 & 34.0 & 1040 & 12334 $\pm$   921 &   315 $\pm$    91 &   4.5 $\pm$  1.3 & 1000 F        &  9898 &  393 &    6 &     16.7  &  536.7  $\pm$  0.4  \\ 
      MSXLMC527 & 10000 & 1020 & 20.0 &  4.0 & 1047 & 11444 $\pm$   173 &   424 $\pm$   486 &  18.5 $\pm$ 10.7 &  886 $\pm$   25 & 10698 &  658 &   19 & $<$  6.9  &                \\ 
     ERO0525406 &  3162 &  356 & 17.8 &  5.6 &  908 &  3873 $\pm$   125 &   220 $\pm$    82 &  23.3 $\pm$  7.1 &  712 $\pm$   46 &  3500 &  280 &   20 &     12.6  &                \\ 
      MSXLMC474 & 11220 & 1447 & 11.2 &  8.2 & 1081 & 12641 $\pm$   169 &   498 $\pm$   571 &  12.5 $\pm$  6.5 &  837 $\pm$   10 & 11910 &  849 &   12 & $<$ 13.5  &  622.8  $\pm$  0.4  \\ 
      IRAS05315 &  5012 &  232 & 20.0 & 11.4 &  901 &  8399 $\pm$    37 &    14 $\pm$     1 &  12.4 $\pm$  1.0 &  266 $\pm$ \phantom{1}3 &  6488 &   57 &   16 &     38.8  &             \\ 
     MSXLMC1780 & 12590 &  737 & 31.6 & 11.5 &  975 & 12857 $\pm$   553 &   243 $\pm$    53 &  18.0 $\pm$  3.4 &  737 $\pm$   29 & 12723 &  423 &   24 &     44.5  &  847.9  $\pm$  0.5  \\ 
      IRAS05373 &  6310 &  844 & 17.8 & 15.9 & 1051 &  7072 $\pm$   118 &   949 $\pm$   166 &  13.0 $\pm$  1.9 & 1163 $\pm$   29 &  6680 &  895 &   15 &     29.3  &  643.7  $\pm$  0.5  \\ 
      MSXLMC971 &  6310 &  863 & 12.6 & 13.4 & 1021 &  6184 $\pm$    53 &   283 $\pm$    14 &   7.6 $\pm$  0.5 &  766 $\pm$   10 &  6247 &  494 &   10 &     26.1  &  648.6  $\pm$  0.3  \\ 
      MSXLMC937 & 15850 & 1250 & 20.0 & 22.9 & 1125 & 16091 $\pm$   203 &   194 $\pm$    17 &   9.0 $\pm$  0.6 &  816 $\pm$ \phantom{1}6 & 15970 &  493 &   13 &     32.1  &  691.0  $\pm$  0.6  \\ 
      IRAS05416 &  7943 &  364 & 28.2 & 12.9 &  941 & 10451 $\pm$    65 &   333 $\pm$    22 &  27.9 $\pm$  3.5 &  879 $\pm$   42 &  9111 &  348 &   28 &     34.5  &  904.0  $\pm$  0.6  \\ 
    LMCLPV76711 &  6310 & 1272 &  4.0 &  9.2 & 1193 &  5988 $\pm$   766 &   486 $\pm$   260 &   5.8 $\pm$  2.5 &  807 $\pm$   84 &  6147 &  786 &    5 &      8.8  &  550.8  $\pm$  0.3  \\ 
      IRAS05495 &  5012 &  355 & 20.0 &  4.0 &  931 & 12277 $\pm$    61 &    12 $\pm$     1 &  20.3 $\pm$  1.2 &  311 $\pm$ \phantom{1}3 &  7844 &   65 &   20 &     57.6  &             \\ 
     ERO0550261 &  7943 &  232 & 35.5 & 16.2 &  884 & 10426 $\pm$    32 &   253 $\pm$    34 &  55.9 $\pm$  4.0 & 1000 F        &  9100 &  242 &   45 &     48.2  & 1111.8  $\pm$  1.2  \\ 
      IRAS05515 &  7943 &  491 & 11.2 & 19.5 & 1006 & 11890 $\pm$   130 &   337 $\pm$    36 &  12.5 $\pm$  1.4 &  963 $\pm$   21 &  9718 &  407 &   12 &     47.5  &  666.1  $\pm$  0.3  \\ 
     MSXLMC1797 &  5012 &  231 & 22.4 & 10.0 &  912 &  5144 $\pm$   266 &   509 $\pm$    90 &  41.3 $\pm$  6.4 & 1000 F        &  5078 &  343 &   30 &     14.3  &  698.8  $\pm$  1.2  \\ 
      IRAS05568 & 10000 &  366 & 35.5 & 14.8 &  922 & 20119 $\pm$   153 &   356 $\pm$    16 &  46.0 $\pm$  3.0 &  798 $\pm$   16 & 14184 &  361 &   40 &     38.9  & 1207.0  $\pm$  0.7  \\ 
      IRAS06028 & 15850 &  809 & 39.8 & 18.6 &  984 & 13708 $\pm$   125 &   218 $\pm$    22 &  14.5 $\pm$  1.4 &  813 $\pm$   15 & 14740 &  420 &   24 &     55.9  &  860.0  $\pm$  0.2  \\ 
      IRAS06108 &  5012 & 1248 & 12.6 & 10.7 & 1041 &  7617 $\pm$    81 &  1483 $\pm$   157 &  17.7 $\pm$  1.4 & 1000 F        &  6179 & 1360 &   15 &     12.6  &  527.5  $\pm$  0.2  \\ 
      IRAS05506 & 14130 &  486 & 25.1 & 24.5 & 1021 & 17934 $\pm$   196 &   134 $\pm$     5 &  28.3 $\pm$  3.3 & 1095 $\pm$   57 & 15919 &  255 &   27 &     28.4  & 1022.4  $\pm$  1.0  \\ 
      IRAS05125 & 14130 &  718 & 39.8 & 12.1 &  968 & 15574 $\pm$    74 &   660 $\pm$    18 &  40.3 $\pm$  1.7 &  869 $\pm$   11 & 14834 &  689 &   40 &      8.1  & 1147.1  $\pm$  1.0  \\ 
     ERO0529379 &  5012 &  233 & 20.0 & 11.4 &  929 &  5638 $\pm$   216 &   430 $\pm$    40 &  32.4 $\pm$  3.2 & 1000 F        &  5316 &  316 &   25 &      7.3  &  685.4  $\pm$  1.1  \\ 
     ERO0518117 &  7943 &  355 & 39.8 &  9.8 &  892 &  9604 $\pm$    22 &   484 $\pm$    58 & 101.4 $\pm$  6.6 &  900 F        &  8734 &  415 &   64 &      7.4  &                \\ 

  \hline
\end{tabular}

\vfill
\end{sidewaystable*}

\end{appendix}

\end{document}